\documentclass[a4paper,11pt]{article}

\pdfoutput=1
\usepackage{pdflscape}
\usepackage{geometry}
\usepackage{jheppub}
\usepackage[utf8]{inputenc}
\usepackage{dsfont}
\usepackage{simplewick}
\usepackage{mathtools,amsmath,amssymb}
\usepackage{graphicx}
\usepackage{lmodern}
\usepackage[T1]{fontenc}
\usepackage{microtype}
\usepackage{hyperref}
\usepackage{xspace}
\usepackage{array}
\usepackage{booktabs}
\usepackage{xspace}
\usepackage{adjustbox}

\makeatletter
\DeclareRobustCommand*{\bfseries}{%
  \not@math@alphabet\bfseries\mathbf
  \fontseries\bfdefault\selectfont
  \boldmath
}

\makeatother

\xspaceaddexceptions{]\}}

\newcommand{\pcl}{\ensuremath{\phi^\text{cl}}\xspace}
\newcommand{\pt}{\ensuremath{\tilde \phi}\xspace}
\newcommand{\ad}{\ensuremath{\mathrm{ad}\xspace}}

\newcommand{\so}[1]{\ensuremath{\mathfrak{so}(#1)\xspace}}
\newcommand{\su}[1]{\ensuremath{\mathfrak{su}(#1)\xspace}}

\newcommand{\gym}{\ensuremath{g_{\mathrm{YM}}\xspace}}
\newcommand{\diff}{\ensuremath{\mathrm{d}\xspace}}

\newcommand{\measy}{\ensuremath{m_{\text{easy}}}}
\newcommand{\measyh}{\ensuremath{\hat{m}_{\text{easy}}}}

\DeclareMathOperator{\tr}{tr}
\def \ph{\phantom}

\newcommand{\bJ}{\ensuremath{\mathbf{J}}\xspace}
\newcommand{\bL}{\ensuremath{\mathbf{L}}\xspace}
\newcommand{\bS}{\ensuremath{\mathbf{S}}\xspace}

\newcommand{\Bfermion}{\ensuremath{\tilde{D}}\xspace}

\title{A Quantum Framework for  AdS/dCFT  through Fuzzy Spherical Harmonics on $S^4$}
\author{Aleix Gimenez-Grau, Charlotte Kristjansen, Matthias Volk and Matthias Wilhelm}

\begin{document}

\begingroup\parindent0pt
\begin{flushright}\footnotesize
DESY 19-220
\end{flushright}
\vspace*{4em}
\centering
\begingroup\LARGE
\bf
A Quantum Framework for  AdS/dCFT  through Fuzzy Spherical Harmonics on $S^4$
\par\endgroup
\vspace{2.5em}
\begingroup\large{\bf Aleix Gimenez-Grau$^{a}$, Charlotte Kristjansen$^{b}$, Matthias Volk$^{b}$ and Matthias Wilhelm$^{b}$}
\par\endgroup
\vspace{1em}
\begingroup\itshape
$^a$DESY Hamburg, Theory Group, \\
Notkestraße 85, 22607 Hamburg, Germany \\[0.5em]

$^b$Niels Bohr Institute, Copenhagen University,\\
Blegdamsvej 17, 2100 Copenhagen \O{}, Denmark

\par\endgroup
\vspace{1em}
\begingroup\ttfamily
aleix.gimenez@desy.de,
kristjan@nbi.ku.dk,
mvolk@nbi.ku.dk,
matthias.wilhelm@nbi.ku.dk \\
\par\endgroup
\vspace{2.5em}
\endgroup

\begin{abstract}

\begin{center}
\bf{Abstract}
\end{center}

\noindent
We consider a non-supersymmetric domain-wall version of ${\cal N} =4$ SYM theory where 
five out of the six scalar fields have non-zero classical values on one side of a wall of codimension one. 
The classical fields have commutators which constitute an irreducible representation 
of the Lie algebra $\so{5}$ leading to a highly non-trivial mixing between color and flavor components of the quantum fields. 
Making use of fuzzy spherical harmonics on $S^4$, we explicitly solve the mixing problem and derive
not only the spectrum of excitations at the quantum level but also the propagators of the original fields needed for perturbative quantum computations. 
As an application, we derive the one-loop one-point function of a chiral primary  and find
complete agreement with a supergravity prediction of the same quantity in a double-scaling limit which involves a limit
of large instanton number in the dual D3-D7 probe-brane setup. 
\noindent
\end{abstract}


\thispagestyle{empty}

\newpage
\hrule
\setcounter{tocdepth}{2}
\tableofcontents
\afterTocSpace
\hrule
\afterTocRuleSpace

\section{Introduction and Summary}
\label{sec:introduction}

There exists a number of domain-wall versions of ${\cal N}=4$ SYM theory characterized by some or possibly all of the scalar fields
acquiring non-vanishing and spacetime-dependent vacuum expectation values (vevs) on one side of a codimension-one 
wall. These theories  constitute defect conformal field theories and  have well-defined holographic duals in the form of probe-brane models with non-vanishing background gauge-field flux or instanton number~\cite{Karch:2000gx,Constable:2001ag,Constable:1999ac,DeWolfe:2001pq,Erdmenger:2002ex,Nagasaki:2012re,Kristjansen:2012tn}. They have been studied both from the perspective of supersymmetric boundary conditions~\cite{Gaiotto:2008sa} and from the perspective of condensed matter physics, the probe-brane models being capable of describing strongly coupled Dirac fermions in
2+1 dimensions~\cite{Myers:2008me,Bergman:2010gm,Grignani:2012jh,Kristjansen:2012ny,Kristjansen:2013hma,Hutchinson:2014lda}. 

More recently, these models have been analyzed from the point of view of integrability, where the domain wall or defect is
viewed as a boundary state of the integrable bulk ${\cal N}=4$ SYM 
theory~\cite{deLeeuw:2015hxa,Buhl-Mortensen:2015gfd,deLeeuw:2016umh,Buhl-Mortensen:2017ind,deLeeuw:2018mkd,deLeeuw:2019sew,Ipsen:2019jne}; see 
also~\cite{Piroli:2017sei,Pozsgay:2018dzs}.
Furthermore, the models have been studied
with the aim of testing AdS/dCFT  in situations where  supersymmetry  is partially or completely 
broken~\cite{Buhl-Mortensen:2016pxs,Buhl-Mortensen:2016jqo,Grau:2018keb},  the comparison between gauge theory and string theory  being made possible by the introduction of a certain double-scaling limit~\cite{Nagasaki:2012re,Kristjansen:2012tn}. Table~\ref{tab:results} below summarizes the status of these investigations.

In the present paper, we fill the last gap in the table. We will study the most complicated of the above mentioned domain-wall versions of ${\cal N}=4$ SYM theory where five out of the six scalar fields have vevs whose commutators
constitute an irreducible representation of the Lie algebra $\so5$.  The string-theory dual of this dCFT  is a D3-D7 probe-brane system where the geometry of the probe brane is $AdS_4\times S^4$, and where a non-Abelian background gauge field forms an instanton bundle with instanton number $d_G$ on  the $S^4$~\cite{Constable:2001ag,Myers:2008me}.
The instanton number on the string-theory side translates into the dimension, $d_G$,
of the $\so5$ representation on the gauge-theory side, where
\begin{equation}
    d_G=\frac{1}{6}(n+1)(n+2)(n+3), \hspace{0.5cm} n\in \mathbb{N}, \label{eq:def-dG}
\end{equation}
Combining the large-$N$ limit with the following double scaling~\cite{Kristjansen:2012tn},
\begin{align}
    \label{eq:double-scaling-limit}
    \lambda \rightarrow \infty, \quad
    n \rightarrow \infty, \quad
    \frac{\lambda}{\pi^2 n^2} \quad \text{fixed},
\end{align}
one can by means of a supergravity approximation derive results for simple observables such as 
one-point functions or  Wilson loops.  
Certain results allow an expansion in positive powers of the double-scaling parameter
$\frac{\lambda}{\pi^2 n^2}$ and open for the possibility of comparing to a perturbative gauge-theory calculation.
We notice that the perturbative regime in the gauge theory lies within the parameter region where the probe-brane system is stable, which is given by~\cite{Myers:2008me}
\begin{equation}
\frac{\lambda}{\pi^2 (n+1)(n+3)}< \frac{2}{7}.
\end{equation}

One simple observable that can be studied using both supergravity and gauge theory is the one-point function of the unique $\so5$-symmetric chiral primary of even length $L$, ${\cal O}_L$.  
In~\cite{Kristjansen:2012tn}, this one-point function was calculated in supergravity to the leading order in the double-scaling parameter. The computation can straightforwardly be extended to subleading order and results in  the following prediction for the ratio between the full one-point function and its tree-level value:
\begin{align}
    \label{eq:sugra-result}
    \frac{\langle \mathcal{O}_L \rangle}{\langle \mathcal{O}_L \rangle_{\text{tree}}}
    =
    1 + \frac{\lambda}{\pi^2 n^2} \frac{L (L + 3)}{4 (L - 1)} + \mathcal{O} \left( \left(\frac{\lambda}{\pi^2 n^2}\right)^2 \right).
\end{align}
This prediction trivially carries over to the simple chiral primary $\tr Z^L$ with $Z=\phi_5+i\phi_6$, which has a non-vanishing projection on the $\so5$-symmetric one. 

In the present paper, we will confirm this supergravity prediction by a rather intricate gauge-theory computation. The non-vanishing $\so5$-symmetric vevs of the scalars introduce a complicated (spacetime-dependent) mass matrix mixing color and
flavor components of the standard fields of ${\cal N}=4$ SYM theory.
Needless to say, the diagonalization of this mass matrix 
requires the machinery of representation theory of orthogonal groups, the key element being the introduction of fuzzy spherical harmonics on $S^4$. 

Our motivation for setting up the perturbative program for this dCFT is not only a wish to reproduce the formula~(\ref{eq:sugra-result}) and thus provide a positive test of AdS/dCFT in a situation where supersymmetry is completely broken.  Having a perturbative program will also make it possible to generate a wealth of new data which could provide input to the boundary conformal
bootstrap program as well as to the search for higher-loop integrability in the one-point function problem in AdS/dCFT.

{%
\renewcommand*{\arraystretch}{1.5}
\renewcommand{\tabcolsep}{1em}
\begin{table}
\begin{center}
\label{tab:results}
\begin{tabular}{l c c c}
    \toprule
 &  D3-D5 & D3-D7 & D3-D7\\ 
     \midrule 
Supersymmetry & 1/2-BPS & None & None  \\
Brane geometry & AdS$_4\times $ S$^2$ & AdS$_4\times $ S$^2$ $\times$ S$^2$& AdS$_4\times $ S$^4$  \\
Flux/Instanton number & $k$ & $k_1, k_2$ & $\frac{(n+1)(n+2)(n+3)} {6}$ \\
Double-scaling parameter   & $\frac{\lambda}{\pi^2k^2}$ & $\frac{\lambda}{\pi^2(k_1^2+k_2^2)}$& $\frac{\lambda}{\pi^2n^2}$  \\
Boundary state & Integrable & Non-integrable & Integrable \\
AdS/dCFT match & Yes &  Yes & Yes (this work) \\
\bottomrule 
\end{tabular}
\end{center}
\caption{The string theory configurations dual to the dCFT versions of ${\cal N}=4$ SYM theory with non-vanishing vevs. The
discussion of the integrability properties of the corresponding boundary states can be found in~\cite{deLeeuw:2018mkd,deLeeuw:2019sew} and the test of the
match between gauge theory and string theory referred to in the first two columns can be found in~\cite{Buhl-Mortensen:2016pxs,Buhl-Mortensen:2016jqo,Grau:2018keb}.}
\end{table}
}
Our paper is organized as follows. We start by describing the diagonalization of the mass matrix in Section~\ref{sec:mass-matrix} and explicitly give the complete spectrum of quantum excitations including their multiplicities.
The propagators of the fields which diagonalize the mass matrix are found following the procedure of~\cite{Buhl-Mortensen:2016jqo}, and due to the spacetime-dependence of the vevs, become propagators in an auxiliary AdS${}_4$ space. For concrete perturbative calculations, it is convenient to have the contraction rules and propagators formulated in terms of the
original fields of ${\cal N}=4$ SYM theory and the complete set of these are presented in Section~\ref{sec:propagators}. In Section~\ref{sec:one-loop-vevs}, we calculate
the one-loop correction to the classical solution as well as to the one-point function of $\tr Z^L$ and confirm the prediction~\eqref{eq:sugra-result} in the double-scaling limit; explicit expression for both quantities at finite $n$ are also attached in an ancillary file to this paper.
Finally, Section~\ref{sec:conclusion} contains our conclusion and outlook. A number of technical
details are relegated to appendices.

\section{Diagonalization of the mass matrix}
\label{sec:mass-matrix}

\subsection{Expansion of the action}

We will be considering a domain-wall version of $\mathcal N = 4$ SYM theory where five of the six real scalar fields $\phi_i$ have
non-vanishing vevs on one side of a codimension-one wall, say for $x_3>0$, and we will be 
interested in calculating observables in this region of spacetime.
With $A_{\mu}^{\text{cl}}=\psi^{\text{cl}}=0$, the classical equations of motion for the six scalars read\footnote{See Appendix~\ref{sec:conventions} for a full set of our conventions. We refer to the 
reviews~\cite{deLeeuw:2017cop,deLeeuw:2019usb}
for an introduction to the study of domain-wall versions of 
${\cal N}=4$ SYM theory and their one-point functions.}
\begin{align}
 \label{eq:classical-eom-sym}
 \nabla^2 \pcl_i = \left[ \pcl_j, \left[ \pcl_j, \pcl_i \right] \right], \quad
 i = 1, \ldots, 6.
\end{align}
A classical solution with \so{5} symmetry was found in~\cite{Constable:2001ag,Castelino:1997rv};
\begin{align}
  \label{eq:solution-eom-so5}
  \phi^{\mathrm{cl}}_i (x)
  &= \frac{1}{\sqrt{2} x_3}
  \begin{pmatrix} G_{i6} & 0 \\  0 & 0 \end{pmatrix}, \quad 
  \pcl_6(x) = 0, \quad x_3> 0.
\end{align}
Here the matrices $G_{i6}$ together with $G_{ij} \equiv -i [G_{i6}, G_{j6}]$ for $i,j = 1, \ldots, 5$ are generators of the representation $(\frac{n}{2},\frac{n}{2},\frac{n}{2})$ of the Lie algebra $\so{6}$.\footnote{We are using the eigenvalues of the three generators of the Cartan subalgebra to label the \so{6} representation, see Appendix~\ref{sec:so5-and-so6}.}
From the commutation relations of $\so{6}$, one can check that (\ref{eq:solution-eom-so5}) indeed solves the equations of motion.
The matrices $G_{i6}$ can be constructed as an $n$-fold symmetrized tensor product of $\gamma$ matrices and their dimension
is given in~\eqref{eq:def-dG}; see Appendix~\ref{sec:g-matrices} for details. 

To take into account quantum effects, we expand the scalar fields around the classical solution~\eqref{eq:solution-eom-so5} as
\begin{align}
 \label{eq:expansion-fields}
 \phi_i(x) = \pcl_i(x) + \pt_i(x).
\end{align}
Inserting the expansion into the action of $\mathcal{N} = 4$ SYM theory generates (spacetime-dependent) mass terms for some of the fields, as well as novel cubic and quartic interaction terms.
This has been worked out in detail in~\cite{Buhl-Mortensen:2016pxs,Buhl-Mortensen:2016jqo,Grau:2018keb}.

Upon insertion of the expansion~\eqref{eq:expansion-fields}, the kinetic terms of the action remain canonical, while the mass terms acquire a non-trivial mixing between different fields.
We can rewrite the mass matrices in a compact form in terms of the operators
\begin{align}
    L_{ij} \equiv \ad \, (G_{ij} \oplus 0_{N-d_G}) \quad i,j=1,\dots,6.
\end{align}
The mass terms split into three different pieces:
\begin{align}
  S_{\text{mass}}
  = S_{\text{m,b,e}}
  + S_{\text{m,b,c}}
  + S_{\text{m,f}}.
\end{align}
The first one involves only bosonic terms, and following~\cite{Buhl-Mortensen:2016jqo} we call it \emph{easy} because the mixing only involves color degrees of freedom,
\begin{align}
\label{eq:easy-action-so5}
  S_{\text{m,b,e}}
  =
  \frac{2}{\gym^2} \int \mathrm{d}^4 x
  \left( \frac{-1}{2 x_3^2} \right) \tr\left[ \frac{1}{2} E^\dag \, \sum_{i=1}^{5} \left(L_{i6}\right)^2 \, E\right], \quad 
  E = 
  \begin{pmatrix}
   A_0 \\
   A_1 \\ 
   A_2 \\
   \pt_6
  \end{pmatrix}.
\end{align}
We call the second term \emph{complicated}, because it mixes color and flavor degrees of freedom,
\begin{align}
  \label{eq:Scomplicated}
 S_{\text{m,b,c}} &=
 \frac{2}{\gym^2} \int \mathrm{d}^4 x
 \left( \frac{-1}{2 x_3^2} \right) \tr \left[ C^\dag
 {%
 \renewcommand*{\arraystretch}{1.5}
 \begin{pmatrix}
  \displaystyle
  \frac{1}{2} \sum_{i=1}^{5} \left(L_{i6}\right)^2 - \frac{1}{2} \sum_{i,j=1}^5 S_{ij} L_{ij} & \; \sqrt{2} \sum_{i=1}^{5}R_i L_{i6} \\
  \sqrt{2} \sum_{i=1}^{5}R_i^\dag L_{i6} & \displaystyle \frac{1}{2} \sum_{i=1}^{5} \left(L_{i6}\right)^2
 \end{pmatrix}
 }
 C \right],
 \end{align}
 with the vector of complicated fields
 \begin{align}
 C &= 
 \begin{pmatrix}
  \pt_1  \\
  \vdots \\
  \pt_5  \\
  A_3
 \end{pmatrix}.
\end{align}
In the above expression, $S_{ij}$ are $5 \times 5$ matrices that form the fundamental representation of $\so{5}$, whereas
 $R_i$ are five-dimensional column vectors with components $(R_j)_k = i \delta_{jk}$.
Finally, we have a mass term for the fermions. 
In this case, not only is there mixing between color and flavor, but also the different chiralities are mixed.
It is therefore useful to separate the fermions into their chiral components using the projectors $P_L = \frac{1}{2}(1+\gamma_5)$ and $P_R = \frac{1}{2}(1-\gamma_5)$.
We obtain
\begin{align}
 \label{eq:mass-term-fermions-inserted-so5}
 \begin{split}
 S_{\mathrm{m,f}}
 &= \frac{2}{\gym^2} \int \mathrm{d}^4 x
 \left( \frac{-1}{2 x_3} \right) \tr \left(
    \bar{\psi}_\alpha  \, \mathcal C_{\alpha\beta}      (P_L \psi_\beta)
  + \bar{\psi}_\alpha  \, \mathcal C_{\alpha\beta}^\dag (P_R \psi_\beta)
 \right).
 \end{split}
\end{align}
The components of $\mathcal C_{\alpha\beta}$ involve the operators $L_{i6}$ and thus act non-trivially on the color part of the fields.
They are explicitly given in Appendix~\ref{sec:details-diagonalization-fermions}.

To set up the perturbative program, we first need to gauge fix introducing ghosts\footnote{For the purpose of diagonalizing the mass matrix, the ghosts behave as easy bosons.} as
 in~\cite{Buhl-Mortensen:2016jqo,Grau:2018keb} and subsequently to
diagonalize the mass matrix, i.e.\ to expand the fields in a basis on which all the operators and matrices in the 
quadratic part of the action act diagonally.
We postpone the somewhat technical construction of this basis to Appendix~\ref{sec:details-diagonalization} and proceed to summarize the spectrum which can largely be understood from the representation theory of \so{5} and \so{6}.

\subsection{Decomposition of the color matrices and easy bosons}

From the color structure of the classical solution~\eqref{eq:solution-eom-so5}, it is natural to decompose the $U(N)$ adjoint fields into blocks as\footnote{The $N \times N$ basis matrices ${F^{m}}_{m'}$ are zero everywhere except at position $(m, m')$, where they are one.}
\begin{align}
 \label{eq:field-decomposition-so5}
 \Phi
 = [\Phi]_{m,m'} {F^m}_{m'}
 + [\Phi]_{m,a}  {F^m}_{a}
 + [\Phi]_{a,m}  {F^a}_{m}
 + [\Phi]_{a,a'} {F^a}_{a'}
 =
 \begin{pmatrix*}[l]
     [\Phi]_{m,m'} & [\Phi]_{m,a} \\
     [\Phi]_{a,m} & [\Phi]_{a,a'}
 \end{pmatrix*},
\end{align}
where $m, m' = 1, \ldots, d_G$ and $a, a' = d_G + 1, \ldots, N$.
Since we rewrote the mass terms using $L_{ij}$, it is natural to ask how it acts on the different blocks.
Anticipating their transformation behavior, we will often refer to $[\Phi]_{m,m'}$ as the fields in the adjoint block, whereas $[\Phi]_{m,a}$ and $[\Phi]_{a,m}$ will simply be called fields in the off-diagonal block.

First, we note that $L_{ij} {F^{a}}_{a'} = 0$, so all the fields in the $(N-d_G)\times(N-d_G)$ block are massless.
We will see in later sections that the fields in this block do not contribute to the one-point functions we will calculate, and we will mostly ignore them.
The fields in the off-diagonal block transform as
\begin{align}
 L_{ij} F^m_{\phantom{m}a} = F^{m'}_{\phantom{m'}a} [G_{ij}]_{m',m}, \quad
 L_{ij} F^a_{\phantom{a}m} = F^{a}_{\phantom{a}m'} [-(G_{ij})^T]_{m',m}.
\end{align}
This means that an upper index $m$ transforms in the $(\tfrac{n}{2}, \tfrac{n}{2}, \tfrac{n}{2})$ of \so{6}, while a lower index $m$ transforms in the dual representation $\overline{(\tfrac{n}{2}, \tfrac{n}{2}, \tfrac{n}{2})}$.
Finally, the fields in the $d_G \times d_G$ dimensional adjoint block carry one index and its dual, so they transform as the product of the two representations.
This product can be decomposed into a direct sum of irreducible representations
\begin{align}
 \label{eq:adj-decomposition-so6}
 \left( \frac{n}{2}, \frac{n}{2}, \frac{n}{2} \right) \otimes
 \overline{\left( \frac{n}{2}, \frac{n}{2}, \frac{n}{2} \right) }
  &= \bigoplus_{m=0}^n (m,m,0).
\end{align}

The key observation (see also~\cite{hep-th/0212170,Steinacker:2015dra}) to obtain the spectrum and diagonalize the easy mass term is that it is given by the difference of Casimir operators for $\so{5}$ and $\so{6}$,
\begin{align}
 \label{eq:so5_easy_term_action}
 \frac{1}{2} \sum_{i=1}^{5} \left(L_{i6}\right)^2
 = \frac{1}{2} \sum_{1\leq i<j\leq 6} (L_{ij})^2 - \frac{1}{2} \sum_{1\leq i<j\leq 5} (L_{ij})^2
 = \frac{1}{2} \left( C_{6} - C_{5} \right).
\end{align}
Any representation of \so{6} can be decomposed into a direct sum of irreducible representations of \so{5}.
Equation~\eqref{eq:so5_easy_term_action} implies that fields belonging to different \so{5} representations will have different masses.

For example, we have seen that the fields in the off-diagonal block transform as the $(\tfrac{n}{2}, \tfrac{n}{2}, \tfrac{n}{2})$ of \so{6} and its dual.
It turns out that they are irreducible representations of \so{5}:
\begin{align}
 \label{eq:decomposition-so5-offdiag}
 [\Phi]_{m,a} 
 :   \left( \frac{n}{2}, \frac{n}{2}, \frac{n}{2} \right)
 \to \left( \frac{n}{2}, 0 \right), \qquad
 [\Phi]_{a,m} 
 :   \overline{\left( \frac{n}{2}, \frac{n}{2}, \frac{n}{2} \right)}
 \to \left( \frac{n}{2}, 0 \right),
\end{align}
where our notation and conventions are explained in Appendix~\ref{sec:so5-and-so6}.
Thus, all fields in the off-diagonal block have the same mass, which we can easily obtain from~\eqref{eq:so5_easy_term_action} and the formulas for the eigenvalues of the Casimirs in~\eqref{eq:casimir-so5-so6-appendix}.

For the adjoint block, we saw in~\eqref{eq:adj-decomposition-so6} that the fields decompose into a sum of irreducible representations of \so{6}. 
Each of these representations of the form $(m, m, 0)$ can in turn be decomposed into \so{5} components using the branching rule~\eqref{eq:ranges-so6}
\begin{align}
 \label{eq:adj-decomposition-so5}
 [\Phi]_{m,m'} : 
 \left( \frac{n}{2}, \frac{n}{2}, \frac{n}{2} \right) \otimes
 \overline{\left( \frac{n}{2}, \frac{n}{2}, \frac{n}{2} \right) }
 \to 
  \bigoplus (L_1, L_2),
\end{align}
where the sum runs over all half-integer $(L_1, L_2)$ such that
\begin{align}
  0 \le L_2 \le L_1, 
  \qquad
  L_1 + L_2 \le n.
\end{align}
Therefore, fields with several different masses occur in the adjoint block, one for each \so{5} representation in the above sum.
Once again, from the expression of the Casimir operators~\eqref{eq:casimir-so5-so6-appendix} we obtain the easy masses summarized in Table~\ref{table:easy_so5_masses}.
It is important to note that the \so{6} Casimir needs to be evaluated for $(L_1+L_2,L_1+L_2,0)$, which can be seen from working out the decomposition~\eqref{eq:adj-decomposition-so5} explicitly.

So far we have only focused on the spectrum, but we have not discussed how the diagonalization can explicitly be carried out.
We can find an explicit orthonormal basis that diagonalizes the easy mass term, namely
\begin{align}
 [\Phi]_{m,m'} F^m_{\ph m m'} 
 = \sum_\bL [\Phi]_{\mathbf L} \hat Y_{\mathbf L}, \qquad
 \tr \, \left(
   \hat Y_{\mathbf L'}^\dag \hat{Y}_{\mathbf L}
 \right)
 = \delta_{\mathbf{L'}, \mathbf{L}}.
\end{align}
The matrices $\hat{Y}_{\bL}$ are \so{5}-symmetric fuzzy spherical harmonics -- the \so{5} analogue of the basis used in~\cite{Buhl-Mortensen:2016pxs,Buhl-Mortensen:2016jqo}.
For our purposes, only the existence of this basis will be important.
An explicit construction of the matrices can be found in~\cite{hep-th/0105006}.
In general, we use the notation $\bL$ to collectively refer to the quantum numbers that uniquely specify an \so{5} state within a representation.
This is described in more detail in Appendix~\ref{sec:so5-and-so6}.
For example, the sum over $\bL$ includes a sum over all possible highest weights $(L_1,L_2)$ in~\eqref{eq:adj-decomposition-so5}, and for each of them also the $d_5(L_1,L_2)$ states that form the representation.

\renewcommand*{\arraystretch}{1.5}
\begin{table}
\centering
\begin{tabular}{c l c}
 \toprule
 Eigenstate &
 \multicolumn{1}{c}{Mass} &
 Multiplicity \\
 \midrule
 $ [E]_{a,a'}$ &
 $ 0 $ &
 $ ( N - d_G ) ( N - d_G ) $ \\
 $ [E]_{m,a}$ &
 $ \measy^2 = \frac{1}{8} n ( n + 4) $ &
 $ 2 d_G ( N - d_G ) $ \\
 $ [E]_{\mathbf L}$ &
 $ \measyh^2 = 2 L_1 L_2 + L_1 + 2 L_2 $ &
 $d_{5}(L_1, L_2)$  \\
 \bottomrule
\end{tabular}
\caption{Masses of the easy bosons $E = A_0, A_1, A_2, \pt_6$. The allowed ranges of $L_1$ and $L_2$ are $0 \le L_2 \le L_1$, $L_1 + L_2 \le n$. The definitions of $d_5$ and $d_G$ can be found in~\eqref{eq:def-dG} and~\eqref{eq:dimension-so5-appendix}, respectively.}
\label{table:easy_so5_masses}
\end{table}
\renewcommand*{\arraystretch}{1.0}

\subsection{Complicated bosons}
\label{sec:complicated-bosons}

We now turn towards the complicated mass terms, for which color and flavor degrees of freedom mix.
The key observation of~\cite{Grau:2018keb} is that if one can find an eigenvector of the $5 \times 5$ block of the mass matrix in~\eqref{eq:Scomplicated} which is annihilated by the $1 \times 5$ block $R_i^\dag L_i$, then we obtain an eigenvector of the full matrix.

In particular, to diagonalize the $5 \times 5$ block we define the total \so{5} `angular momentum' operator $J_{ij}$, such that
\begin{align}
 \label{eq:cg-decompositon-so5}
 J_{ij} \equiv L_{ij} + S_{ij} 
 \quad \Rightarrow \quad
 \frac{1}{2} \sum_{i,j=1}^5 S_{ij} L_{ij}
 = \frac{1}{2} \sum_{1 \leq i < j \leq 5} \Big[ (J_{ij})^2 - (L_{ij})^2 - (S_{ij})^2 \Big].
\end{align}
On the right hand side, we have a combination of \so{5} Casimir operators, which act trivially on irreducible representations.
As mentioned above, the matrices $S_{ij}$ form the fundamental of \so{5} which is labeled by $(\tfrac{1}{2}, \tfrac{1}{2})$.
After decomposing the fields in \so{5} fuzzy spherical harmonics, they therefore transform in the product representation $(L_1, L_2) \otimes (\tfrac{1}{2}, \tfrac{1}{2})$.
This product decomposes into irreducible representations with well-defined total angular momentum $(J_1, J_2)$ as
\begin{subequations}
\label{eq:5_times_irrep}
\begin{align}
 \label{eq:5_times_irrep_general}
 \begin{split}
 (L_1, L_2) \otimes (\tfrac{1}{2}, \tfrac{1}{2} ) &= 
 (L_1 + \tfrac{1}{2}, L_2 + \tfrac{1}{2}) \oplus
 (L_1 - \tfrac{1}{2}, L_2 - \tfrac{1}{2}) \oplus
 (L_1, L_2) \\
 &\quad \oplus
 (L_1 + \tfrac{1}{2}, L_2 - \tfrac{1}{2}) \oplus
 (L_1 - \tfrac{1}{2}, L_2 + \tfrac{1}{2}), \quad\text{for }0<L_2<L_1, 
 \end{split} \\
 \label{eq:5_times_irrep_funny_1}
 (L_1, L_1) \otimes (\tfrac{1}{2}, \tfrac{1}{2} ) &= 
 (L_1 + \tfrac{1}{2}, L_1 + \tfrac{1}{2}) \oplus
 (L_1 - \tfrac{1}{2}, L_1 - \tfrac{1}{2}) \oplus
 (L_1 + \tfrac{1}{2}, L_1 - \tfrac{1}{2}),  \\
 \label{eq:5_times_irrep_funny_2}
 (L_1, 0) \otimes (\tfrac{1}{2}, \tfrac{1}{2} ) &= 
 (L_1 + \tfrac{1}{2}, \tfrac{1}{2}) \oplus
 (L_1, 0) \oplus
 (L_1 - \tfrac{1}{2}, \tfrac{1}{2}).
\end{align}
\end{subequations}
The masses of the fields that diagonalize the $5 \times 5$ block of the complicated action can now again be obtained from the Casimir operators,
\begin{align}
 \label{eq:eigenvalue-5x5}
 \frac{1}{2} \bigg( \sum_{i=1}^{5} \left(L_{i6}\right)^2 - \sum_{i,j=1}^5 S_{ij} L_{ij} \bigg)
 =
 \frac{1}{2} \left[ 
     C_{6}(L_1 + L_2, L_1 + L_2, 0) -
     C_{5}(J_1, J_2)
     + C_{5}\left(\tfrac{1}{2}, \tfrac{1}{2} \right) \right].
\end{align}

Generically, we obtain the five fields $B_{\pm, \pm}$, $B_{\pm, \mp}$ and $B_{00}$ from the decomposition~\eqref{eq:5_times_irrep_general} that diagonalize the $5 \times 5$ block.
It turns out that $B_{\pm, \pm}$ and $B_{00}$ are indeed mass eigenstates of the full complicated mass term, as the corresponding basis states are annihilated by $\sum_{i=1}^5R^{\dagger}_i L_{i6}$.
As we describe in Appendix~\ref{sec:details-diagonalization-bosons}, the remaining complicated fields $B_{\pm, \mp}$ and $A_3$ still mix through a $3 \times 3$ matrix.
Diagonalizing this matrix we find the six mass eigenstates $B_{\pm, \pm}$, $B_{00}$, $D_{\pm}$ and $D_{0}$, where the last three are simple linear combinations of $B_{\pm, \mp}$ and $A_3$.
We list their masses in Table~\ref{tab:masses-6x6}.
There are two edge cases in the decomposition of $(L_1, L_2) \otimes (\tfrac{1}{2}, \tfrac{1}{2})$ corresponding to \eqref{eq:5_times_irrep_funny_1} and \eqref{eq:5_times_irrep_funny_2}.
We find that
    for $(L_1, L_1)$ the $B_{00}$ and $D_0$ fields are missing, and
    for $(L_1, 0)$ the $B_{--}$ and $D_0$ fields are missing.
This concludes the derivation of the spectrum for the complicated bosons in the adjoint block.

\renewcommand*{\arraystretch}{1.5}
\begin{table}
  \centering
  \begin{adjustbox}{center}
  \begin{tabular}{c l c}
    \toprule
    Eigenstate & 
    \multicolumn{1}{c}{Mass} &
    Multiplicity \\
    \midrule
    $ B_{++} $
    & $\hat m_{++}^2 = (2 L_1 + 1) L_2$
    & $d_{5}(L_1+\tfrac12,L_2+\tfrac12)$ \\
    $ B_{--} $
    & $\hat m_{--}^2 = (2 L_1 + 3) (L_2 + 1)$ 
    & $d_{5}(L_1-\tfrac12,L_2-\tfrac12)$ \\
    $ B_{00} $
    & $\hat m_{00}^2 = L_1 + 2 L_2 (L_1 + 1) + 2$ 
    & $d_{5}(L_1,L_2)$ \\
    $ D_0 $
    & $\hat m_{0}^2 = L_1 + 2 L_2 (L_1 + 1) + 2$ 
    & $d_{5}(L_1,L_2)$ \\
    $ D_{+} $
    & $\hat m_{+}^2 = 1 + (L_1 + 2 L_2 (L_1 + 1)) 
      + \sqrt{1 + 4(L_1 + 2 L_2 (L_1 + 1))}$ 
    & $d_{5}(L_1,L_2)$ \\
    $ D_{-} $
    & $\hat m_{-}^2 = 1 + (L_1 + 2 L_2 (L_1 + 1)) 
      - \sqrt{1 + 4(L_1 + 2 L_2 (L_1 + 1))}$ 
    & $d_{5}(L_1,L_2)$ \\
    \bottomrule
  \end{tabular}
  \end{adjustbox}
  \caption{Masses and eigenstates of the complicated bosons in the adjoint block. The allowed ranges of $L_1$ and $L_2$ are $0 \le L_2 \le L_1$, $L_1 + L_2 \le n$. Note that in the case $L_2 = L_1$ the $B_{00}$ and $D_0$ fields are missing, and in the case $L_2 = 0$ the $B_{--}$ and $D_0$ fields are missing.
  The definition of $d_5$ can be found in~\eqref{eq:dimension-so5-appendix}.}
  \label{tab:masses-6x6}
\end{table}
\renewcommand*{\arraystretch}{1.0}

The diagonalization for the complicated bosons in the off-diagonal block proceeds in a similar manner.
In this case, the relevant decomposition is
\begin{align}
 \left( \tfrac{n}{2}, 0 \right) \otimes \left( \tfrac{1}{2}, \tfrac{1}{2} \right)
 = \left( \tfrac{n+1}{2}, \tfrac{1}{2} \right)
 \oplus \left( \tfrac{n}{2}, 0\right)
 \oplus \left( \tfrac{n-1}{2}, \tfrac{1}{2} \right).
\end{align}
In this case, $B_{00}$ and $A_3$ mix in a $2 \times 2$ matrix which is diagonalized by $D_{\pm}$.
We list the spectrum of the fields in the off-diagonal blocks in Table~\ref{tab:masses-6x6-offdiagonal}.
By abuse of notation, we reuse some of the previous names for the diagonal fields.

\renewcommand*{\arraystretch}{1.5}
\begin{table}
  \centering
  \begin{tabular}{c l c}
    \toprule
    Eigenstate & 
    \multicolumn{1}{c}{Mass} &
    Multiplicity \\
    \midrule
    $ B_{++} $
    & $m^2_{++} = \frac{1}{8} n^2$
    & $2 d_{5}(\tfrac{n+1}{2}, \tfrac{1}{2}) (N - d_G)$ \\
    $ B_{-+} $
    & $m^2_{-+} =\frac{1}{8} (n+4)^2$
    & $2 d_{5}(\tfrac{n-1}{2}, \tfrac{1}{2}) (N - d_G)$ \\
    $ D_{+} $
    & $m^2_{+} =\frac{1}{8} \left(n^2 + 4n + 8 + 4 \sqrt{2(n^2 + 4n + 2)}\right)$
    & $2 d_{5}(\tfrac{n}{2}, 0) (N - d_G)$ \\
    $ D_{-} $
    & $m^2_{-} =\frac{1}{8} \left(n^2 + 4n + 8 - 4 \sqrt{2(n^2 + 4n + 2)}\right)$
    & $2 d_{5}(\tfrac{n}{2}, 0) (N - d_G)$ \\
    \bottomrule
  \end{tabular}
  \caption{Masses and eigenstates of the complicated bosons in the off-diagonal block. The definition of $d_5$ can be found in~\eqref{eq:dimension-so5-appendix}.}
  \label{tab:masses-6x6-offdiagonal}
\end{table}
\renewcommand*{\arraystretch}{1.0}

\subsection{Fermions}

The diagonalization of the fermionic mass matrix $\mathcal C_{\alpha\beta}$ is non-trivial, so we will consider first a simplified version of the problem.
The observation we make is that the eigenvalues of $\mathcal C^\dag \mathcal C$ are actually the fermionic masses squared. 
Moreover, we will use the eigenvectors of $\mathcal C^\dag \mathcal C$ to construct the eigenvectors of $\mathcal C$.
From the explicit form of $\mathcal C_{\alpha\beta}$ given in Appendix~\ref{sec:details-diagonalization-fermions}, we obtain
\begin{align}
 \mathcal C^\dag \mathcal C 
 = \frac{1}{2} \bigg( \sum_{i=1}^{5} \left(L_{i6}\right)^2 - \sum_{i,j=1}^5 \tilde S_{ij} L_{ij} \bigg).
\end{align}
The $4 \times 4$ matrices $(\tilde{S}_{ij})_{\alpha\beta}$ constitute the four-dimensional representation of \so{5} which is labelled by $(\frac{1}{2}, 0)$.

Notice the similarity of this problem with that of the $5 \times 5$ block of the complicated bosonic mass term.
In particular, a variant of~\eqref{eq:eigenvalue-5x5} still holds, with the difference that now the total angular momentum $(J_1,J_2)$ takes values in the decomposition\footnote{We also have to change the last term in~\eqref{eq:eigenvalue-5x5} to $C_{5}(\tfrac{1}{2}, 0)$.}
\begin{subequations}
 \begin{align}
 \label{eq:decomposition-fermions-so5}
  (L_1, L_2) \otimes (\tfrac{1}{2}, 0)
  & = (L_1 + \tfrac{1}{2}, L_2) \oplus (L_1 - \tfrac{1}{2}, L_2) \oplus
      (L_1, L_2 + \tfrac{1}{2}) \oplus (L_1, L_2 - \tfrac{1}{2})\,,\\
 \nonumber &\hspace{0.5\textwidth}     \text{ for }0<L_2<L_2,\\
  \label{eq:decomposition-fermions-so5-special1}
  (L_1, L_1) \otimes (\tfrac{1}{2}, 0)
  & = (L_1 + \tfrac{1}{2}, L_1) \oplus (L_1, L_1 - \tfrac{1}{2})\,, \\
  \label{eq:decomposition-fermions-so5-special2}
  (L_1, 0) \otimes (\tfrac{1}{2}, 0)
  & = (L_1 + \tfrac{1}{2}, 0) \oplus (L_1 - \tfrac{1}{2}, 0) \oplus
      (L_1, \tfrac{1}{2})\,.
\end{align}
\end{subequations}
It is now an easy exercise to extract the masses of the fermionic diagonal fields.
Note that compared to the complicated bosons there is no further mixing of fields after coupling the \so{5} representations $(L_1, L_2)$ and $(\tfrac{1}{2}, 0)$ appropriately.
In analogy to the previous section, we will denote the diagonal fields by $\Bfermion_{\alpha, \beta}$.
The fermionic masses are listed in Table~\ref{tab:fermion-masses-so5} for the adjoint block and in Table~\ref{tab:fermion-masses-so5-offdiagonal} for the off-diagonal block.

\renewcommand*{\arraystretch}{1.5}
\begin{table}
  \centering
  \begin{tabular}[h]{c l c}
    \toprule
    Eigenstate & 
    \multicolumn{1}{c}{Mass} & 
    Multiplicity \\
    \midrule
    $\Bfermion_{+0}$ & 
    $\hat m_{+0} = \sqrt{2(L_1 + 1) L_2}$ &
    $d_{5}(L_1 + \tfrac{1}{2}, L_2)$ \\
    $\Bfermion_{-0}$ & 
    $\hat m_{-0} = \sqrt{2 (L_1 + 1) (L_2 + 1)}$ &
    $d_{5}(L_1 - \tfrac{1}{2}, L_2)$ \\
    $\Bfermion_{0+}$ & 
    $\hat m_{0+} = \sqrt{\frac{1}{2}(2 L_1 + 1) (2 L_2 + 1)}$ &
    $d_{5}(L_1, L_2 + \tfrac{1}{2})$ \\
    $\Bfermion_{0-}$ & 
    $\hat m_{0-} =\sqrt{\frac{1}{2}(2 L_1 + 3)(2 L_2 + 1)}$ &
    $d_{5}(L_1, L_2 - \tfrac{1}{2})$ \\
    \bottomrule
  \end{tabular}
  \caption[Spectrum of the fermions (adjoint block)]{Mass eigenvalues of the fermions in the adjoint block.  The allowed ranges of $L_1$ and $L_2$ are $0 \le L_2 \le L_1$, $L_1 + L_2 \le n$. Note that in the case $L_2 = L_1$ the fields $\Bfermion_{-0}$ and $\Bfermion_{0+}$ are missing, and in the case $L_2 = 0$ the $\Bfermion_{0-}$ fields are missing.
  The definition of $d_5$ can be found in~\eqref{eq:dimension-so5-appendix}.}
  \label{tab:fermion-masses-so5}
\end{table}
\renewcommand*{\arraystretch}{1.0}

\renewcommand*{\arraystretch}{1.5}
\begin{table}
  \centering
  \begin{tabular}[h]{c l c}
    \toprule
    Eigenstate & 
    Mass &
    Multiplicity \\
    \midrule
    $\Bfermion_{+0}$ &
    $m_{+0} = \frac{1}{\sqrt{8}}n$ &
    $2 d_{5}(\tfrac{n+1}{2}, 0) (N - d_G )$ \\
    $\Bfermion_{-0}$ & 
    $m_{-0} = \frac{1}{\sqrt{8}}(n+4)$ &
    $2 d_{5}(\tfrac{n-1}{2}, 0) (N - d_G )$\\
    $\Bfermion_{0+}$ & 
    $m_{0+} = \frac{1}{\sqrt{8}}(n+2)$ &
    $2 d_{5}(\tfrac{n}{2}, \tfrac{1}{2}) (N - d_G )$ \\
    \bottomrule
  \end{tabular}
  \caption[]{Mass eigenvalues of the fermions in the off-diagonal block. The definition of $d_G$ can be found in~\eqref{eq:def-dG}.}
  \label{tab:fermion-masses-so5-offdiagonal}
\end{table}
\renewcommand*{\arraystretch}{1.0}

\section{Propagators}
\label{sec:propagators}

In the previous section, we have presented the spectrum of `masses' of all the fields in the theory.
In the action, these masses combine with a spacetime-dependent factor into
 $\tfrac{m^2}{x_3^2}$ for the bosons, and $\tfrac{m}{x_3}$ for the fermions.
The propagators of fields in $(d + 1)$-dimensional Minkowski space with such spacetime-dependent mass terms are related to the propagators of fields in $AdS_{d + 1}$, as observed 
in~\cite{Nagasaki:2011ue,Buhl-Mortensen:2016pxs,Buhl-Mortensen:2016jqo}.

For the purpose of our computation in Section \ref{sec:one-loop-vevs}, only the propagators of fields evaluated at the same point in spacetime will be relevant.
Since they are divergent, we need to introduce a regulator to keep them finite, and we will accomplish this working in dimensional reduction with $d = 3 - 2 \epsilon$, such that the codimension of the defect remains one.
For the bosonic fields, the regulated propagator is~\cite{Buhl-Mortensen:2016jqo}
\begin{align}
    \label{eq:spacetime-propagator-regularized}
    \begin{split}
        K^{m^2}(x, x)
        &= \frac{\gym^2}{2} \frac{1}{16 \pi^2 x_3^2} \bigg[ m^2 \bigg(\!\!-\frac{1}{\epsilon} - \log(4 \pi) + \gamma_{\mathrm{E}} - 2 \log(x_3) + 2 \Psi( \nu + \tfrac{1}{2} ) - 1 \bigg) - 1 \bigg],
    \end{split}
\end{align}
where $\nu = \sqrt{m^2 + \tfrac{1}{4}}$.
Similarly, the (spinor trace of the) regularized propagator for the fermions is
\begin{align}
    \label{eq:fermion-prop-regularized}
        \tr K_F^{m}(x,x)
        = \frac{\gym^2}{8 \pi^2 x_3^3}
        \Bigg[ &
        m^3 + m^2 - 3m - 1 \\
        &  + m(m^2 - 1) \left( -\frac{1}{\epsilon} - \log(4 \pi) + \gamma_{\mathrm{E}} - 2 \log(x_3) + 2 \Psi(m) - 2 \right)
        \Bigg].
        \nonumber
\end{align}
In the above expressions, $\Psi(x)$ is the digamma function and $\gamma_E$ is the Euler-Mascheroni constant.

As discussed in Section~\ref{sec:mass-matrix}, one can change basis from the fields $\pt_i$, $A_\mu$ and $\psi_\alpha$ in the action to the diagonal fields $B_{\pm, \pm}$, $B_{00}$, $D_{\pm}$ and $D_{0}$, such that the mass terms become diagonal.
The propagators between these diagonal fields are then of the form~\eqref{eq:spacetime-propagator-regularized} and~\eqref{eq:fermion-prop-regularized} we just presented.
However, it is easier to perform field-theory computations if we know the propagators between the original fields in the action.
This can be achieved by inverting the steps in the diagonalization procedure, as explained in more detail in~\cite{Buhl-Mortensen:2016jqo,Grau:2018keb}. 
In the resulting propagators there is mixing between color and flavor degrees of freedom, which is introduced by the presence of matrix elements of $\so{6}$ generators.

Throughout this section, we denote by $K^{m_i^2}$ the scalar propagator with the mass $m_i^2$ being one of the masses listed in Tables~\ref{tab:masses-6x6}-\ref{tab:fermion-masses-so5-offdiagonal}, and similarly for the fermions.
We will merely present the final results in the main text and refer the reader to Appendix~\ref{sec:propagator-details} for more details.

\subsection{Off-diagonal block}
\label{sec:prop-off-diagonal-block}

We begin with the propagators between fields from the off-diagonal block, because they are the most important ones for the purposes of later calculations in the large-$N$ limit.
We remind the reader that these fields are of the form $[\Phi]_{m,a}$, where $m = 1, \ldots, d_G$ and $a = d_G + 1, \ldots, N$.
The propagators will be expressed in terms of the matrix elements $[G_{ij}]_{m,m'}$ of the matrices $G_{ij}$ that appear in the classical solution; see Appendix~\ref{sec:g-matrices} for more details.

The simplest propagator is the one between two easy fields $E = A_0, A_1, A_2, \pt_6$, because in this case there is no mixing between the flavor and the color structure,
\begin{align}
  \langle [E]_{m,a} [E]_{m',a'}^\dagger \rangle 
  = \delta_{m,m'} \delta_{a,a'} K^{\measy^2}.
\end{align}
Note that the propagator between two different easy fields vanishes.

The remaining scalars $\pt_i$ with $i = 1, \ldots, 5$ mix with each other in the following way:
\begin{align}
    \label{eq:prop-bosons-complicated-offdiagonal}
    \begin{split}
        \langle [\pt_i]_{m,a} [\pt_j]_{m',a'}^\dagger \rangle
        &=
        \delta_{a,a'} \Bigg[
        \delta_{ij} \delta_{m,m'} f^{\mathrm{sing}}
        + [G_{ij}]_{m,m'}         f^{\mathrm{lin}}
        + 4 [G_{i6} G_{j6}]_{m,m'}        f^{\mathrm{prod}} \Bigg].
    \end{split}
\end{align}
The functions $f$ above are linear combinations of bosonic propagators, with coefficients which only depend on $n$:
\begin{align}
    \label{eq:def-fs-bosons-offdiagonal}
    \begin{split}
        f^{\mathrm{sing}}
        &= \frac{n}{2 (n + 2)} K^{m_{-+}^2} + \frac{n + 4}{2 (n + 2)} K^{m_{++}^2}, \\
        f^{\mathrm{lin}}
        &= \frac{i}{n + 2} \left( K^{m_{-+}^2} - K^{m_{++}^2} \right), \\
        f^{\mathrm{prod}}
        &= - \frac{K^{m_{-+}^2}}{2 n (n + 2)} - \frac{K^{m_{++}^2}}{2 (n + 2) (n + 4)} + \frac{K^{m_-^2}}{4 N_+} + \frac{K^{m_+^2}}{4 N_-},
    \end{split}
\end{align}
where the (normalization) factor $N_\pm$ is given by
\begin{align}
    \label{eq:definition-normalization-factors}
    N_\pm = 4 \measy^2 + 1 \pm \sqrt{4 \measy^2 + 1}\,.
\end{align}

As discussed in the diagonalization, the five scalars $\pt_i$ and the third component of the gauge field also couple in a non-trivial way,
\begin{align}
    \langle [\pt_i]_{m,a} [A_3]_{m',a'}^\dagger \rangle
    =
    -i \delta_{a,a'} \frac{1}{\sqrt{n (n + 4) + 2}} [G_{i6}]_{m,m'} \left( K^{m_-^2} - K^{m_+^2} \right),
\end{align}
while the third component of the gauge field with itself gives
\begin{align}
    \langle [A_3]_{m,a} [A_3]_{m',a'}^\dagger \rangle
    =
    \frac{\delta_{a,a'} \delta_{m,m'}}{2} \left[
        \left(1 + \frac{1}{\sqrt{4 \measy^2 + 1}}\right) K^{m_-^2} +
        \left(1 - \frac{1}{\sqrt{4 \measy^2 + 1}}\right) K^{m_+^2}
    \right].
\end{align}
Note the similarity between these propagators, and the ones obtained for the defect theory dual to a D3-D7 setup with $\so{3} \times \so{3} $ symmetry~\cite{Grau:2018keb}. 
In that case, the propagators had precisely the same structure if one makes the schematic replacement $G_{i6} \to t_i$, where $t_i$ are generators of $\so{3} \times \so{3} $ (see (3.25)-(3.29) of~\cite{Grau:2018keb} for further details).

Finally, in the diagonalization of the fermions $\psi_\alpha$ with $\alpha = 1, \ldots, 4$, different chiralities are mixed with the color and flavor degrees of freedom.
As a result, the propagators will contain $\gamma_5$. 
Moreover, matrix elements $(C_i)_{\alpha\beta}$ will appear, where $C_i$ are the matrices that couple scalars and fermions in the action of $\mathcal N = 4$ SYM theory, see~\eqref{eq:def-dimred-g-matrices}.
The propagators have the following structure:\footnote{The notation $[jkl]$ denotes antisymmetrization of the three indices, normalized by $\tfrac{1}{3!}$.}
\begin{align}
    \label{eq:ferm-prop-offdiag}
    \begin{split}
        \langle [\psi_\alpha]_{m,a} & [\overline{\psi_\beta}]_{a',m'} \rangle
        =  
        \delta_{a,a'} \Bigg[        
        \delta_{m,m'} \left[
        \delta_{\alpha,\beta} 
        \left(f^{0,+}_{F} - \gamma_5 f^{0,+}_{F} \gamma_5 \right)
        + i (C_6)_{\alpha,\beta} 
        \left(f^{0,-}_{F} + \gamma_5 f^{0,-}_{F} \gamma_5 \right)
        \right] \\
        & - \delta_{\alpha\beta} [G_{45}]_{m,m'} 
        \left( f^{1,+}_{F} + \gamma_5 f^{1,+}_{F} \gamma_5 \right)
        + i \sum_{i=1}^3 \sum_{j=4}^5 (C_i C_j)_{\alpha,\beta} [G_{ij}]_{m,m'}
        \left(\gamma_5 f_{F}^{1,-} - f_{F}^{1,-} \gamma_5 \right) \\
        & + \Bigg( 
          \frac{1}{2} \sum_{i,j,k=1}^3 \epsilon_{ijk} (C_i)_{\alpha,\beta}  [G_{jk}]_{m,m'}
          + i (C_6)_{\alpha,\beta}  [G_{45}]_{m,m'}
        \Bigg)
        \left(f_{F}^{1,-} - \gamma_5 f_{F}^{1,-} \gamma_5 \right) \\
        & + \sum_{i=1}^3 (C_{i})_{\alpha,\beta} [G_{i6}]_{m,m'} 
          \left( f_{F}^{2,+} + \gamma_5 f_{F}^{2,+} \gamma_5 \right)
        - \sum_{i=4}^5 (C_{i})_{\alpha,\beta} [G_{i6}]_{m,m'}
          \left(\gamma_5 f_{F}^{2,+} + f_{F}^{2,+} \gamma_5 \right) \\
        & + i \sum_{i=1}^3 (C_{i}C_6)_{\alpha,\beta} [G_{i6}]_{m,m'} 
          \left( f_{F}^{2,-} - \gamma_5 f_{F}^{2,-} \gamma_5 \right) \\
        & - i \sum_{i=4}^5 (C_{i}C_6)_{\alpha,\beta} [G_{i6}]_{m,m'}
          \left(\gamma_5 f_{F}^{2,-} - f_{F}^{2,-} \gamma_5 \right) \\
        & + \frac{i}{2} \sum_{i,j,k=1}^3 \sum_{l=4}^5 \epsilon_{ijk} (C_i C_l)_{\alpha,\beta} 
        [G_{[j6} G_{k6} G_{l6]}]_{m,m'}
        \left( \gamma_5 f_{F}^{3} + f_{F}^{3} \gamma_5 \right) \\
        & + \frac{1}{2} \sum_{i,j,k=4}^6 \sum_{l=1}^3 \epsilon_{ijk} (C_i C_l)_{\alpha,\beta} 
        [G_{[j6} G_{k6} G_{l6]}]_{m,m'}
        \left( f_{F}^{3} + \gamma_5 f_{F}^{3} \gamma_5 \right)
        \Bigg].
    \end{split}
\end{align}
As for the complicated bosons, the $f_F$ are functions that depend on $n$ and the fermionic propagators~\eqref{eq:fermion-prop-regularized}
\begin{align}
    \begin{split}
        f^{0,\pm}_{F}
        &=
        \frac{(n+4)}{8 (n+1)} K_F^{m_{+0}}
        \pm \frac{n (n+4)}{4 (n+1) (n+3)} K_F^{m_{0+}}
        + \frac{n }{8 (n+3)} K_F^{m_{-0}}, \\
        f_F^{1,\pm}
        &=
        \pm \frac{1}{4 (n+1)} K_F^{m_{+0}}
        + \frac{1}{2 (n+1) (n+3)} K_F^{m_{0+}}
        \mp \frac{1}{4 (n+3)} K_F^{m_{-0}}, \\
        f_F^{2,\pm}
        &=
            \frac{1}{4 (n+1)} K_F^{m_{+0}}
        \pm \frac{(n+2)}{2(n+1) (n+3)} K_F^{m_{0+}}
        +   \frac{1}{4 (n+3)} K_F^{m_{-0}}, \\
        f_F^{3}
        & =
        - \frac{3}{(n+1) (n+2)} K_F^{m_{+0}}
        - \frac{6}{(n+1) (n+2) (n+3)} K_F^{m_{0+}}
        + \frac{3}{(n+2) (n+3)} K_F^{m_{-0}}.
    \end{split}
\end{align}
The fermionic masses $m_{\alpha\beta}$ can be found in Table~\ref{tab:fermion-masses-so5-offdiagonal}.

\subsection{Adjoint block}
\label{sec:prop-central-block}

Now we present the propagators in the adjoint block.
In this case, the fields are $[\Phi]_{m,m'}$ with $m = 1, \ldots, d_G$, but it is convenient to express them in terms of irreducible $\so{5}$ representations. 
As explained in Section~\ref{sec:mass-matrix}, this is achieved by changing basis: $[\Phi]_{m,m'} F^m_{\ph mm'} = [\Phi]_\bL \hat{Y}_\bL$.
In particular, the matrix elements of generators $L_{ij} = \ad \, G_{ij}$ will appear, and they can be computed as
\begin{align}
    \langle \bL | L_{ij} | \bL' \rangle = \tr \left( \hat{Y}_\bL^\dagger L_{ij} \hat{Y}_{\bL'} \right)
    = \tr \left( \hat{Y}_\bL^\dagger \left[ G_{ij}, \hat{Y}_{\bL'} \right] \right).
\end{align}
However, using this expression is hard in general, because we do not have explicit formulas for $\hat{Y}_\bL$.
What one can do instead is to compute the matrix elements thinking of $L_{ij}$ as an operator acting on an abstract vector $|\mathbf L \rangle$ in a certain $\so{5}$ representation.
We give a prescription on how to do this in Appendix~\ref{sec:action-generators}.

The propagator between two easy fields $E = A_0, A_1, A_3, \pt_6$ is simple because there is no mixing of color and flavor
\begin{align}
    \langle [E]^{\ph{\dagger}}_{\bL} [E]_{\bL'}^\dagger \rangle 
    = \delta_{\bL, \bL'} K^{\measyh^2}.
\end{align}
For the propagators between the five scalars $\pt_i$ with $i = 1, \ldots, 5$, the resulting structure is more complicated than in the off-diagonal block:
\begin{align}
    \label{eq:prop-bosons-complicated}
    \begin{split}
        \langle [\pt_i]^{\ph{\dagger}}_{\bL} [\pt_j]_{\bL'}^\dagger \rangle
        &=
        \delta_{ij} \delta_{\bL, \bL'} \, \hat f^{\,\mathrm{sing}}
        + \langle \bL | L_{ij} | \bL' \rangle \, \hat f^{\,\mathrm{lin}}
        + \langle \bL | \{ L_{ik}, L_{jl} \} L_{kl} | \bL' \rangle \, 
        \hat f^{\,\mathrm{cubic}}\\
        & \quad
        + \langle \bL | \{ L_{ik}, L_{kj} \} | \bL' \rangle \, \hat f^{\,\mathrm{sym}}_{5} \\
        & \quad
        + \langle \bL | \{ L_{i6}, L_{6j} \} | \bL' \rangle
        \left[
            \delta_{L_1, L_1'} \delta_{L_2, L_2'} \, \hat f^{\,\mathrm{sym}}_{6}
            + \delta_{L_1', L_1 \pm 1} \delta_{L_2', L_2 \mp 1} \, 
            \hat f^{\,\mathrm{opp}}
        \right],
    \end{split}
\end{align}
and
\begin{align}
    \begin{aligned}
        \langle [\pt_i]^{\ph{\dagger}}_{\bL} [A_3]^\dag_{\bL'} \rangle
        = i \langle \bL | L_{i6} | \bL' \rangle
        ( \delta_{L_1, L_1' + \frac{1}{2}} \delta_{L_2, L_2' - \frac{1}{2}}
        + \delta_{L_1, L_1' - \frac{1}{2}} \delta_{L_2, L_2' + \frac{1}{2}})
        \hat f^{\,\phi A}(L_1', L_2').
    \end{aligned}
\end{align}
The third component of the gauge field has the following propagator:
\begin{align}
    \label{eq:propagator-A3-A3-main}
    \langle [A_3]^{\ph{\dagger}}_{\bL} [A_3]_{\bL'}^\dagger \rangle &=
    \delta_{\bL, \bL'} \left(
        \frac{\left(-1 + \sqrt{4 \measyh^2 + 1}\right)^2}{2 N_-} K^{\hat m_+^2} +
        \frac{\left(1 + \sqrt{4 \measyh^2 + 1}\right)^2}{2 N_+} K^{\hat m_-^2}
    \right),
\end{align}
where $N_\pm$ were introduced in~\eqref{eq:definition-normalization-factors}.\footnote{Note that $N_\pm$ needs to be evaluated using $\measyh^2$ instead of $\measy^2$.}

Finally, one can obtain the propagators between the fermions in the adjoint block in a similar manner.
Rewriting the propagators in terms of matrix elements is a complex task, and in most applications only certain traces of them will appear.
In particular, one has that
\begin{align}
        \tr \, \langle [\psi_\alpha]_{\bL} [\overline{\psi_\beta}]_{\bL'} \rangle
        & = \sum_{i=1}^3 (C_{i})_{\alpha,\beta} 
            \langle \bL | L_{i6} | \bL' \rangle
            \tr \hat f_{F}^{\mathrm{lin}}(L_1,L_2;L_1',L_2') \nonumber\\
        & \phantom{{}={}}+ \Bigg( 
        \sum_{i,j,k=4}^6 \sum_{l=1}^3
        \epsilon_{ijk} (C_i C_l)_{\alpha,\beta} \,
        \langle \bL | L_{[j6} L_{k6} L_{l6]} | \bL' \rangle 
    \label{eq:ferm-prop-main}\\ \nonumber
        & \phantom{{}={}} \hphantom{ + \Bigg(} - \frac{i}{3}
        \sum_{i,j,k,l=1}^3
        \epsilon_{ijk} (C_i C_l)_{\alpha,\beta} \,
        \langle \bL | L_{[j6} L_{k6} L_{l6]} | \bL' \rangle
        \Bigg) \tr \hat f_{F}^{\mathrm{cub}}(L_1,L_2;L_1',L_2'),
\end{align}
and
\begin{align}
    \label{eq:ferm-prop-main-g5}
        \tr \Big( 
          \gamma_5 \langle [\psi_\alpha]_{\bL} [\overline{\psi_\beta}]_{\bL'} & \rangle
        \Big)
        = - \sum_{i=4}^5 (C_{i})_{\alpha,\beta}
            \langle \bL | L_{i6} | \bL' \rangle
            \tr \hat f_{F}^{\mathrm{lin}}(L_1,L_2;L_1',L_2') \\\nonumber
        & + i \sum_{i,j,k=1}^3 \sum_{l=4}^6
         \epsilon_{i j k} (C_{i} C_{l})_{\alpha,\beta} \,
        \langle \bL | L_{[j6} L_{k6} L_{l6]} | \bL' \rangle
        \tr \hat f_{F}^{\mathrm{cub}}(L_1,L_2;L_1',L_2').
\end{align}
The full propagators $\langle [\psi_\alpha]_{\bL} [\overline{\psi_\beta}]_{\bL'} \rangle$ would have a structure similar to that of~\eqref{eq:ferm-prop-offdiag}, but containing many more terms and matrix elements of products of generators $L_{ij}$ up to cubic order.

As for the off-diagonal case, the functions $\hat{f}_F$ are linear combinations of the propagators between mass eigenstates~\eqref{eq:spacetime-propagator-regularized} and~\eqref{eq:fermion-prop-regularized}.
Again, these functions only depend on the labels $(L_1,L_2)$ of the external fields.
However, since their expressions are more involved than in the off-diagonal case, we postpone their explicit formulas until Appendix~\ref{sec:propagator-details}.

\section{One-loop corrections to the classical solution and one-point functions} 
\label{sec:one-loop-vevs}

Following previous work~\cite{Buhl-Mortensen:2016jqo,Grau:2018keb}, we will now use the propagators to compute the first quantum correction to the vacuum expectation value of the five scalars $\phi_i$ for $i = 1, \ldots 5$, as well as the one-loop one-point function of the 1/2-BPS operator $\tr(Z^L)$, where $Z = \phi_5 + i \phi_6$.
Throughout this section we will work in the large-$N$ limit, and we will specify which results are applicable for finite $n$ or in the large-$n$ regime.\footnote{It should also be possible to extend this to finite $N$ following \cite{Buhl-Mortensen:2016jqo,Guo_2017}.}
One-loop corrections to one-point functions of more general, non-protected operators can similarly be obtained in analogy with \cite{Buhl-Mortensen:2016jqo,Grau:2018keb}.

\subsection{One-loop correction to the classical solution}
\label{sec:correction-classical-solution}

The first quantum correction to the classical solution is given by the contraction of an external scalar with an effective three-vertex,
\begin{align}
    \label{eq:one-loop-corr-vevs}
    \contraction{%
        \langle \phi_i \rangle_{\text{1-loop}}(x)
        =
    }{\pt_i}{(x)
        \int \text{d}^4 y
        \sum_{\Phi_1, \Phi_2, \Phi_3}
        V_3(}{\Phi}
    \contraction{%
        \langle \phi_i \rangle_{\text{1-loop}}(x)
        =
        \pt_i(x)
        \int \text{d}^4 y
        \sum_{\Phi_1, \Phi_2, \Phi_3}
        V_3(\Phi_1(y),}{\Phi}{_2(y), }{\Phi}
    \langle \phi_i \rangle_{\text{1-loop}}(x)
    =
    \pt_i(x)
    \int \text{d}^4 y
    \sum_{\Phi_1, \Phi_2, \Phi_3}
    V_3(\Phi_1(y), \Phi_2(y), \Phi_3(y)).
\end{align}
The sum on the right-hand side runs over all fields in the theory. We show in Appendix~\ref{sec:effective-vertex} that
\begin{align}
 \sum_{\Phi_1, \Phi_2, \Phi_3} V_3
 \contraction{(\Phi_1(y), }{\Phi_2}{(y), }{\Phi_3}
 (\Phi_1(y), \Phi_2(y), \Phi_3(y))
 = - \frac{4 \sqrt{2} N}{\pi^2 (y_3)^3} W(n) \tr \left(\pt_i G_{i6} \right).
\end{align}
The function $W(n)$ is positive for $n \ge 0$, and is given explicitly
\begin{align}
\label{eq:effective-vertex}
\begin{split}
 W(n) =
 -\frac{1}{64} \Bigg(
 & \frac{2 (n-4) (n+8)}{n (n+4)}
   + \frac{2 \sqrt{2} (n+2) (n (n+4)-4)}{(n+1) (n+3)}
   + \frac{n \left(n^2-8\right) \Psi (m_{+0})}{2 (n+1)} \\
 & + \frac{(n+2)^2 (n (n+4)-4) \Psi (m_{0+})}{(n+1) (n+3)}
   + \frac{(n+4) (n (n+8)+8) \Psi (m_{-0})}{2 (n+3)} \\
 & - \frac{\left(n^4+8 n^3-32 n+8 n^2+64\right) 
     \Psi \big(\nu_{\text{easy}} + \frac{1}{2}\big)}{n (n+4)} \\
 & -\frac{n^3 (n+5) \Psi \big(\nu_{++} + \frac{1}{2}\big)}{2 (n+2) (n+4)}
   - \frac{(n-1) (n+4)^3 \Psi \big(\nu_{-+} +\frac{1}{2}\big)}{2 n (n+2)}
 \Bigg),
\end{split}
\end{align}
in terms of the masses of bosons and fermions in the off-diagonal blocks (see Tables~\ref{tab:masses-6x6-offdiagonal} and~\ref{tab:fermion-masses-so5-offdiagonal}) and $\nu_i = \sqrt{m_i^2 + \tfrac{1}{4}}$.
We also attach a completely explicit expression for $W(n)$ in an ancillary file to this paper.
In Section~\ref{sec:correction-tr-ZM} we will be interested in this function in the double-scaling limit~\eqref{eq:double-scaling-limit}.
Expanding for $n \rightarrow \infty$, this function simplifies dramatically:
\begin{align}
\label{eq: W expansion}
 W(n) = \frac{1}{4n^2} + \mathcal O(n^{-3})\,.
\end{align}
From the individual terms in~\eqref{eq:effective-vertex}, one would expect terms growing as fast as $n^2 \log(n)$ in the large-$n$ limit.
However, from the supergravity calculation we know that all terms growing faster than $1/n^2$ should not be present.
This ``miraculous'' cancellation provides a very non-trivial check for our results.

Moreover, using the relation between the matrices $G_{i6}$ and the \so{5} fuzzy spherical harmonics given in Appendix~\ref{sec:g-matrices}, we can compute the contraction
\begin{align} 
 \contraction{}{\pt_i}{\tr \Big(}{\pt_j}
 \pt_i \tr \Big(\pt_j G_{j6} \Big) 
 &= K^{m^2 = 6}(x, y) \, G_{i6}.
\end{align}
The remaining spacetime integral was already computed in~\cite{Grau:2018keb}:
\begin{align}
\label{eq:Veff-spacetime-integral}
 \int \diff^4 y \frac{1}{y_3^3} K^{m^2 = 6}(x, y)
 =\frac{\gym^2}{2} \frac{1}{4 x_3}.
\end{align}

Assembling the pieces, we see that the one-loop correction to the classical solution is proportional to the classical solution such that we can write
\begin{align}
    \label{eq:phi-oneloop}
 \langle \phi_i(x) \rangle
 = \left( 
   1 
   - \frac{\lambda}{\pi^2} W(n) 
   + \mathcal O \left( \lambda^2 \right)
 \right) \langle \phi_i(x) \rangle_{\text{tree}}.
\end{align}
We note that this correction is non-vanishing, fitting the picture observed so far that for a domain-wall setup which conserves part of the supersymmetry there is no correction to the classical field~\cite{Buhl-Mortensen:2016jqo}  whereas for setups which break the supersymmetry there can be a correction~\cite{Grau:2018keb}.
The one-loop corrections to vanishing classical vevs are all vanishing.

\subsection{One-loop correction to \texorpdfstring{$\langle\tr(Z^L)\rangle$}{<tr(Z**L)>}}
\label{sec:correction-tr-ZM}

\begin{figure}[!tbp]
    \centering
    \begin{minipage}[b]{0.99\textwidth}
        \def\svgwidth{\textwidth}
\begingroup%
  \makeatletter%
  \providecommand\color[2][]{%
    \errmessage{(Inkscape) Color is used for the text in Inkscape, but the package 'color.sty' is not loaded}%
    \renewcommand\color[2][]{}%
  }%
  \providecommand\transparent[1]{%
    \errmessage{(Inkscape) Transparency is used (non-zero) for the text in Inkscape, but the package 'transparent.sty' is not loaded}%
    \renewcommand\transparent[1]{}%
  }%
  \providecommand\rotatebox[2]{#2}%
  \newcommand*\fsize{\dimexpr\f@size pt\relax}%
  \newcommand*\lineheight[1]{\fontsize{\fsize}{#1\fsize}\selectfont}%
  \ifx\svgwidth\undefined%
    \setlength{\unitlength}{461.56181525bp}%
    \ifx\svgscale\undefined%
      \relax%
    \else%
      \setlength{\unitlength}{\unitlength * \real{\svgscale}}%
    \fi%
  \else%
    \setlength{\unitlength}{\svgwidth}%
  \fi%
  \global\let\svgwidth\undefined%
  \global\let\svgscale\undefined%
  \makeatother%
  \begin{picture}(1,0.33306682)%
    \lineheight{1}%
    \setlength\tabcolsep{0pt}%
    \put(0,0){\includegraphics[width=\unitlength,page=1]{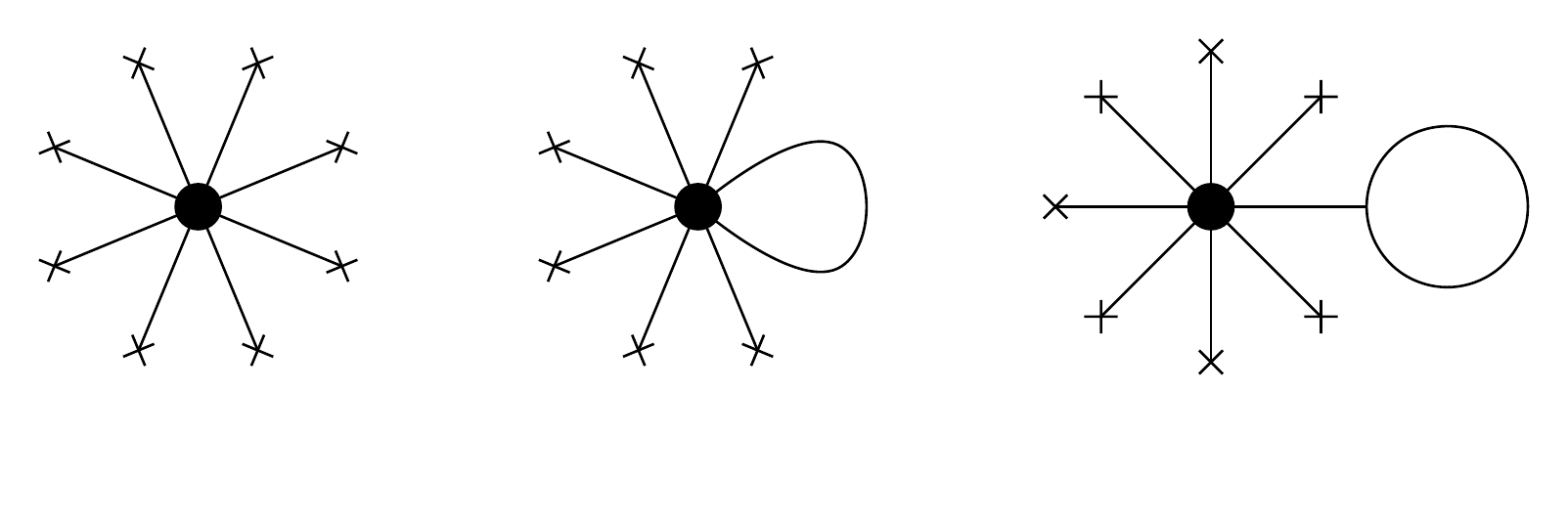}}%
    \put(0.05616155,0.02742199){\color[rgb]{0,0,0}\makebox(0,0)[lt]{\lineheight{1.25}\smash{\begin{tabular}[t]{l}(a) Tree level\end{tabular}}}}%
    \put(0.38783832,0.0295219){\color[rgb]{0,0,0}\makebox(0,0)[lt]{\lineheight{1.25}\smash{\begin{tabular}[t]{l}(b) Tadpole\end{tabular}}}}%
    \put(0.76020815,0.0295219){\color[rgb]{0,0,0}\makebox(0,0)[lt]{\lineheight{1.25}\smash{\begin{tabular}[t]{l}(c) Lollipop\end{tabular}}}}%
  \end{picture}%
\endgroup%

    \end{minipage}
    \caption{Diagrams (identical to the ones of~\cite{Buhl-Mortensen:2016jqo})  that contribute at tree level (a) and one-loop order (b)-(c) to a single-trace operator such as $\langle \tr Z^L \rangle_{L=8}$ (in the planar limit).
        The black dot denotes the operator and the crosses signify the insertion of the classical solution.
          }
    \label{fig:feynman-diagrams-one-loop-order}
\end{figure}
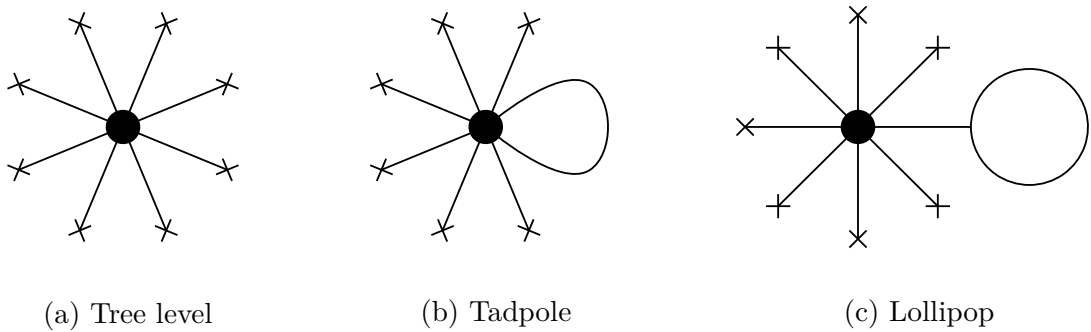

Next, we consider the scalar single-trace operator $\tr(Z^L)$ with $Z = \phi_5 + i \phi_6$ and aim to compute the first quantum correction to its one-point function.

At tree level, the one-point function $\langle\tr(Z^L)\rangle$ was first computed in~\cite{deLeeuw:2016ofj}; it is simply obtained by inserting the classical solution $Z^{\mathrm{cl}} = \phi_5^{\text{cl}}$ into the trace:
\begin{align}
    \label{eq:trZL-tree}
    \langle \tr Z^L \rangle_{\text{tree}} = \frac{1}{(\sqrt{2}x_3)^L}\tr G_{56}^L=
    \begin{cases}
        0, & L \quad \text{odd}, \\
        \frac{1}{(\sqrt{2} x_3)^L} \left[
            \frac{2}{L+3} B_{L+3}(-\tfrac{n}{2}) 
          - \frac{(n+2)^2}{2(L+1)} B_{L+1}(-\tfrac{n}{2}) 
        \right], & L \quad \text{even},
    \end{cases}
\end{align}
where $B_l$ denotes the $l$-th Bernoulli polynomial.

The general procedure for computing the one-loop one-point function of scalar single-trace operators can be found in~\cite{Buhl-Mortensen:2016pxs,Buhl-Mortensen:2016jqo,Grau:2018keb}.
As was derived there, there are only two contributions for the operator $\tr(Z^L)$, which were called tadpole and lollipop, see Figure~\ref{fig:feynman-diagrams-one-loop-order}:
\begin{equation}
 \langle \tr Z^L \rangle_{\mathrm{1-loop}}=\langle \tr Z^L \rangle_{\mathrm{tad}}+\langle \tr Z^L \rangle_{\mathrm{lol}}.
\end{equation}
In particular, since the operator is 1/2-BPS, there is no correction to its wave function as well as no renormalization.

The tadpole diagram corresponds to inserting the classical solution for $L - 2$ scalars and contracting the remaining two fields.
This can be done in $L$ inequivalent ways, so we obtain
\begin{align}
    \langle \tr Z^L \rangle_{\mathrm{tad}} = L \tr \left[
        (Z^{\mathrm{cl}})^{L-2}
        \contraction{}{Z}{\;}{Z}
        Z \; Z
    \right].
\end{align}
The contraction of $Z$ with itself is simply
\begin{align}
    \contraction{}{Z}{\;}{Z}
    Z \; Z =
    \contraction{}{\pt}{_5\;}{\pt}
    \pt_5 \; \pt_5 -
    \contraction{}{\pt}{_6\;}{\pt}
    \pt_6 \; \pt_6\,,
\end{align}
since $\phi_5$ and $\phi_6$ are an easy and a complicated field respectively and there is no propagator that mixes them.
Using the propagators presented in the previous section and taking into account that only the fields in the off-diagonal block contribute in the large-$N$ limit, we find
\begin{align}
    \label{eq:tadpole-non-expanded}
    \begin{split}
        \langle \tr Z^L \rangle_{\mathrm{tad}} =
        L N \Bigg[ &
        \tr \left( (Z^{\mathrm{cl}})^{L - 2} \right) \left( f^{\text{sing}} - K^{m_{\text{easy}}^2} \right)
        + 4 \tr \left( (Z^{\mathrm{cl}})^{L - 2} G_{56} G_{56} \right) f^{\text{prod}}
        \Bigg],
    \end{split}
\end{align}
using the combinations of propagators $f$ given in~\eqref{eq:def-fs-bosons-offdiagonal}.
Note that this gives us the contribution of the tadpole for any finite value of $n$, because the color trace is known in terms of Bernoulli polynomials, see~\eqref{eq:trZL-tree}.
As for the effective vertex $W(n)$, we have a cancellation of the regulator-dependent terms coming from the spacetime propagator for any finite $n$.

In order to compare our result to the supergravity prediction, we need to evaluate the expression in the large-$n$ limit.
Inserting the expression for the traces~\eqref{eq:trZL-tree} into~\eqref{eq:tadpole-non-expanded} and expanding for $n \rightarrow \infty$, we find that the leading order term is
\begin{align}
    \label{eq:trZL-tad-final}
    \langle \tr Z^L \rangle_{\mathrm{tad}}
    \xrightarrow{n \rightarrow \infty}
    \frac{\lambda}{\pi ^2 n^2}
    \frac{L (L+1)}{2 (L-1)}
    \langle \tr Z^L \rangle_{\mathrm{tree}}.
\end{align}
Notice how once again, only terms which are at most of order $n^{-2}$ contribute in the large-$n$ limit, even though from~\eqref{eq:tadpole-non-expanded} one could expect a growth-rate faster than this.

The second type of diagram is the lollipop diagram, which is nothing but the one-loop correction to the classical solution for one of the scalars in the operator.
We find, using our result~\eqref{eq:phi-oneloop},
\begin{align}
    \label{eq:trZL-lol-final}
 \langle \tr Z^L \rangle_{\mathrm{lol}} 
 = L \tr \left[
     (Z^{\mathrm{cl}})^{L-1}
     \langle Z \rangle_{\text{1-loop}}
 \right]
 = - \frac{\lambda L}{\pi^2} W(n) \langle \tr Z^L \rangle_{\mathrm{tree}}
 \xrightarrow{n \rightarrow \infty}
 - \frac{\lambda L}{4 \pi^2 n^2} \langle \tr Z^L \rangle_{\mathrm{tree}}.
\end{align}
In the last step, we have used the expansion \eqref{eq: W expansion} of $W(n)$ for $n \rightarrow \infty$.

Combining the tree-level result~\eqref{eq:trZL-tree} with the values of the tadpole and lollipop diagrams~\eqref{eq:trZL-tad-final} and~\eqref{eq:trZL-lol-final} respectively, we find
\begin{align}
    \frac{\langle \tr Z^L \rangle}{\langle \tr Z^L \rangle_{\text{tree}}}
    =
    1 + \frac{\lambda}{\pi^2 n^2} \frac{L (L + 3)}{4 (L - 1)} + \mathcal{O} \left( \left(\frac{\lambda}{\pi^2 n^2}\right)^2 \right).
\end{align}
Up to first order in the double-scaling parameter, this matches precisely the result from the supergravity computation~\eqref{eq:sugra-result}.
Note that as in~\cite{Buhl-Mortensen:2016jqo,Grau:2018keb} we are actually forced to consider the above ratio in order to compare the supergravity to the field-theory result:
the supergravity result computes the one-point function of the unique $\so{5}$-symmetric
chiral primary on which the operator $\tr(Z^L)$ has a non-vanishing projection.

A completely explicit expression for $\langle \tr Z^L \rangle_{\mathrm{1-loop}}$ at finite $n$ is attached in an ancillary file to this paper.

\newpage
\section{Conclusion and Outlook}

Making use of fuzzy spherical harmonics on $S^4$, we have set up the framework required to carry out perturbative calculations
of observables in the domain-wall version of ${\cal N}=4$ SYM theory where five scalar fields have $\so{5}$-symmetric vevs in a half-space. As an application, we have computed the one-loop correction to the one-point function
of a specific chiral primary and found that it agrees in a double-scaling limit with the prediction from a supergravity computation in the dual string-theory setup. We notice that
a match between gauge and string theory is obtained for all defect setups of the given type regardless of whether supersymmetry is fully or only partially broken and regardless
of whether the relevant boundary state is characterized as integrable or non-integrable, cf.\ Table~\ref{tab:results}.

With the perturbative framework fully developed, one can of course compute other types of observables of the dCFT, such as more general correlation functions or Wilson loops.  The study of Wilson loops in the closely related dCFT dual to the D3-D5 probe-brane system listed in Table~\ref{tab:results}  has revealed interesting novel examples of Gross-Ooguri like
phase transitions~\cite{Nagasaki:2011ue,deLeeuw:2016vgp,Aguilera-Damia:2016bqv,Preti:2017fhw,Bonansea:2019rxh}.  Furthermore, the investigation of two-point functions
in the same setup has led to new insights concerning conformal data of dCFTs~\cite{deLeeuw:2017dkd,Widen:2017uwh} and in general such data might prove useful as input for 
the boundary conformal bootstrap program~\cite{Billo:2016cpy,Liendo:2016ymz,Mazac:2018biw}.

 The one-loop contribution to the one-point function of general non-protected operators in the present $\so{5}$-symmetric setup could potentially provide important information for  the integrability program. The corresponding boundary state  has been argued to be integrable~\cite{deLeeuw:2018mkd} and the derivation of a closed formula for all tree-level one-point
functions is in progress~\cite{deLeeuw:2020}. 
Explicit results at one-loop order might make it possible to package the results for the two leading orders into one formula, put forward a proposal for
an asymptotic formula for higher loop orders as was done for the D3-D5 case~\cite{Buhl-Mortensen:2017ind} and eventually bootstrap an exact all-loop order formula for both cases. 

From the string-theory perspective, the most burning open problem is to understand the reason for the integrability or non-integrability of the boundary states associated
with the different probe-brane models considered here, cf.\ Table~\ref{tab:results}. 

\label{sec:conclusion}

\section*{Acknowledgments}
C.K., M.V. and M.W. were supported in part by DFF-FNU through the grant DFF-FNU 4002-00037.
M.V. and M.W.\ were moreover supported by the ERC starting grant 757978 and the research grant 00015369 from Villum Fonden.
M.W. was also supported by the research grant 00025445 from Villum Fonden.
A.G. was partially supported by the DFG through project number 400570283.
A.G would like to thank the NBI for hospitality during the completion of this work.

\appendix

\section{Conventions}
\label{sec:conventions}

\subsection{\texorpdfstring{$\mathcal N = 4$}{N=4} SYM action}
\label{sec:sym-action}

Throughout our work, we consider a mostly-positive metric $\eta^{\mu \nu} = \mathrm{diag}(-1, +1, +1, +1)$.
The action of $\mathcal N = 4$ SYM theory is given by
\begin{align}
 \label{eq:action-n4-sym}
  S_{\mathcal N = 4} = \frac{2}{\gym^2} \int \diff^4 x \, \tr \Bigg(
& - \frac{1}{4} F_{\mu \nu} F^{\mu \nu}
  - \frac{1}{2} D_{\mu} \phi_i D^{\mu} \phi_i
  + \frac{i}{2} \bar \psi \gamma^\mu D_\mu \psi \\ \nonumber
& + \frac{1}{4} [ \phi_i, \phi_j ] [ \phi_i, \phi_j ]
  + \frac{1}{2} \sum_{i = 1}^3 \bar \psi C_i [ \phi_i, \psi]
  + \frac{1}{2} \sum_{i = 4}^6 \bar \psi C_i [ \phi_i, \gamma_5 \psi]
 \Bigg),
\end{align}
where $(C_i)_{\alpha\beta}$ are $4\times 4$ matrices of Clebsch-Gordan coefficients that couple the two spinors with the scalars. 
We will use the same conventions as~\cite{Buhl-Mortensen:2016jqo}:
\begin{align}
\label{eq:def-dimred-g-matrices}
\begin{alignedat}{3}
 & C_1 \equiv C^{(1)}_1 = i
 \begin{pmatrix}
 0 & -\sigma_3 \\ \sigma_3 & 0
 \end{pmatrix},
 \quad
 && C_2 \equiv C^{(1)}_2 = i
 \begin{pmatrix}
  0 & \sigma_1 \\ -\sigma_1 & 0
 \end{pmatrix},
 \quad
 && C_3 \equiv C^{(1)}_3 =
 \begin{pmatrix}
  \sigma_2 & 0 \\ 0 & \sigma_2
 \end{pmatrix},
 \\
 & C_4 \equiv C^{(2)}_1 = i
 \begin{pmatrix}
  0 & -\sigma_2 \\ -\sigma_2 & 0
 \end{pmatrix},
 \quad
 && C_5 \equiv C^{(2)}_2 =
 \begin{pmatrix}
  0 & -\mathds{1}_2 \\ \mathds{1}_2 & 0
 \end{pmatrix},
 \quad
 && C_6 \equiv C^{(2)}_3 =  i
 \begin{pmatrix}
  \sigma_2 & 0 \\ 0 & -\sigma_2
 \end{pmatrix}.
\end{alignedat}
\end{align}
The matrices in the first line are Hermitian, $(C^{(1)}_i)^\dagger = C^{(1)}_i$, while those in the second are anti-Hermitian, $(C^{(2)}_i)^\dagger = -C^{(2)}_i$.
Furthermore, we note some useful properties:
\begin{alignat}{2}
 &  \left\{ C^{(1)}_i, C^{(1)}_j \right\} = + 2 \delta_{ij}, \quad
 && \left\{ C^{(2)}_i, C^{(2)}_j \right\} = - 2 \delta_{ij}, \\
 &  \left[  C^{(1)}_i, C^{(1)}_j \right] 
    = - 2 i \epsilon_{ijk} C^{(1)}_k, \quad
 && \left[  C^{(2)}_i, C^{(2)}_j \right] = - 2 \epsilon_{ijk} C^{(2)}_k,
\end{alignat}
and the two sets commute $\left[ C^{(1)}_i, C^{(2)}_j \right] = 0$.

\subsection{\texorpdfstring{\so{5}}{so(5)} and \texorpdfstring{\so{6}}{so(6)}}
\label{sec:so5-and-so6}

Given an $\so{n}$ Lie Algebra, we normalize the generators $L_{ij} = -L_{ji}$ such that
\begin{align}
 \label{eq:comm_rels_son}
 [ L_{ij}, L_{kl} ]
 &= i \left( 
   \delta_{ik} L_{jl}
 + \delta_{jl} L_{ik}
 - \delta_{jk} L_{il} 
 - \delta_{il} L_{jk} \right)
 \quad \mathrm{for} \quad 
 i,j,k,l = 1, \ldots, n.
\end{align}

We will label our representations in terms of the quantum numbers of the highest weight.
Our conventions follow~\cite{hecht1965} since we will make use of some of the Clebsch-Gordan coefficients for coupling different \so{5} representations published there.
For $\so{5}$, we need two quantum numbers $(L_1,L_2)$ to specify a representation, which correspond to the eigenvalues of $\frac{1}{2}(L_{12} \pm L_{34})$ acting on the highest weight state.
The most relevant examples for our work will be
\begin{align}
 \so{5} \,\, : \quad
 \mathbf{4} = (\tfrac{1}{2},0), \quad
 \mathbf{5} = (\tfrac{1}{2},\tfrac{1}{2}), \quad
 \mathbf{10} = (1,0).
\end{align}
Our notation is related to the $\so{5}$ Dynkin labels (e.g.\ used in~\cite{Feger:2012bs}) by $(L_1,L_2) = [2 L_2, 2 (L_1-L_2)]$.

Similarly, for $\so{6}$ we need three quantum numbers $(P_1,P_2,P_3)$, which correspond to the eigenvalues of $L_{12}$, $L_{34}$ and $L_{56}$ acting on the highest weight state.
Some simple examples are
\begin{align}
 \so{6} \,\, : \quad
 \mathbf{4}       = (\tfrac{1}{2},\tfrac{1}{2},\tfrac{1}{2}), \quad
 \bar{\mathbf{4}} = (\tfrac{1}{2},\tfrac{1}{2},-\tfrac{1}{2}).
\end{align}
Our notation is related to the $\so{6}$ Dynkin labels by $(P_1,P_2,P_3) = [P_1-P_2,P_2+P_3,P_2-P_3]$.
With our conventions, the dimensions of the irreducible $\so{5}$ and $\so{6}$ representations are
\begin{align}
  \label{eq:dimension-so5-appendix}
  d_{5} \left(L_1, L_2 \right) 
& = \frac{1}{6} 
  (2 L_1 + 2 L_2 + 3)
  (2 L_1 - 2 L_2 + 1)
  (2 L_2 + 1)
  (2 L_1 + 2), \\
  \label{eq:dimension-so6}
  \begin{split}
  d_{6} (P_1, P_2, P_3) 
& = \frac{1}{12}
    (1 + P_1 - P_2) (3 + P_1 + P_2) (2 + P_1 - P_3) \times \\
  &~\hspace{2.3em}
    (1 + P_2 - P_3) (2 + P_1 + P_3) (1 + P_2 + P_3).
  \end{split}
\end{align}

The Casimir operator is defined as the sum over all independent generators squared:
\begin{align}
 \label{eq:casimir_son}
 C_n = \sum_{i < j} (L_{ij})^2.
\end{align}
With our normalizations, it has eigenvalues
\begin{subequations}
 \label{eq:casimir-so5-so6-appendix}
 \begin{align}
 \label{eq:casimir-so5-appendix}
 C_{5}(L_1,L_2) & = 2 \Big[ L_1(L_1 + 2) + L_2 (L_2 + 1) \Big], \\[0.5em]
 \label{eq:casimir-so6-appendix}
 C_{6} (P_1, P_2, P_3) & = P_1 (P_1 + 4) + P_2 ( P_2 + 2) + P_3^2.
\end{align}
\end{subequations}

Let us also write the branching rule of $\so{6}$ representations into $\so{5}$,
\begin{align}
 \label{eq:ranges-so6}
 (P_1,P_2,P_3) \to \bigoplus (L_1,L_2)\,, \quad \text{where} \quad
 P_3 \le L_1 - L_2 \le P_2 \le L_1 + L_2 \le P_1.
\end{align}
The most relevant cases for us are $(P_1, P_2, P_3) = (\tfrac{n}{2}, \tfrac{n}{2}, \tfrac{n}{2})$ which implies $(L_1, L_2) = (\tfrac{n}{2}, 0)$ for the fields in the off-diagonal block, and $(P_1, P_2, P_3) = (L_1 + L_2, L_1 + L_2, 0)$ for the fields in the adjoint block.

To label the states in a given \so{5} representation, we use the collective label $\bL = (L_1, L_2) \, \ell_1 \ell_2 \, m_{1} m_{2}$.
Here $m_1$ and $m_2$ are the eigenvalues of the two Cartan generators $\tfrac{1}{2} (L_{12} + L_{34})$ and $\tfrac{1}{2}(L_{12} - L_{34})$ covering the ranges $m_i = -\ell_i, \ldots, +\ell_i$.
The spins $\ell_i$ are subject to the constraints
\begin{align}
    -L_1 + L_2 \leq \ell_1 - \ell_2 \leq L_1 - L_2 \leq \ell_1 + \ell_2 \leq L_1 + L_2,
\end{align}
and $\ell_1 + \ell_2 \in \mathbb{Z}$~\cite{hecht1965}.

\subsection{\texorpdfstring{$G$}{G} matrices}
\label{sec:g-matrices}

Consider a four-dimensional representation of the \so5 Clifford algebra
\begin{align}
 \label{eq:clifford-algebra}
 \{\gamma_i, \gamma_j \} = 2 \delta_{ij} \mathds{1}_{4 \times 4}.
\end{align}
This can be used as a building block for some particular types of \so5 and \so6 representations as follows.
Take the $n$-fold tensor product and project to $\mathrm{Sym}(\otimes^n \mathbb C^4)$ as
\begin{align}
 G_{i6} = \frac{1}{2}
 \big( \underbrace{\gamma_i \otimes 1 \otimes \cdots \otimes 1}_{n \; \text{factors}} + \cdots + 1 \otimes \cdots \otimes 1 \otimes \gamma_i \big)_{\mathrm{sym}},
\end{align}
and define
\begin{align}
 \label{eq:def_g_matrices_so6}
 G_{ij} \equiv -i \left[G_{i6}, G_{j6}\right], \quad i, j = 1, \ldots, 5.
\end{align}
From the anticommutation relations~\eqref{eq:clifford-algebra}, one can verify that $G_{ij}$ for $i,j = 1, \ldots, 5$ satisfy the commutation relations of \so5 and $G_{ij}$ for $i,j = 1, \ldots, 6$ satisfy the commutation relations of \so6.
We also refer to the appendix of~\cite{Castelino:1997rv}, where some useful identities for the matrices $G_{ij}$ can be found.
The matrices $G_{i6}$ are related to the \so{5} fuzzy spherical harmonics $\hat{Y}_{\mathbf{J}}$ by
\begin{equation}
\begin{aligned}
    G_{16} &= \frac{1}{\sqrt{2}} a_n \left( \hat{Y}_{++} + \hat{Y}_{--} \right), \quad
    &&G_{26} = -\frac{i}{\sqrt{2}} a_n \left( \hat{Y}_{++} - \hat{Y}_{--} \right), \\
    G_{36} &= - \frac{1}{\sqrt{2}} a_n \left( \hat{Y}_{-+} - \hat{Y}_{+-} \right), \quad
    &&G_{46} = -\frac{i}{\sqrt{2}} a_n \left( \hat{Y}_{-+} + \hat{Y}_{+-} \right), \\
    G_{56} &= - a_n \hat{Y}_{00},
\end{aligned}
\end{equation}
where
\begin{align}
 a_n = \frac{1}{2} \sqrt{\frac{1}{5} \, n(n + 4) \, d_{5}(\tfrac{n}{2}, 0)},
 \quad \text{and} \quad
 \hat{Y}_{\alpha \beta} \equiv \hat{Y}_{(\frac{1}{2}, \frac{1}{2}) \frac{1}{2} \frac{1}{2} \alpha \beta}, \;
 \hat{Y}_{0 0} \equiv \hat{Y}_{(\frac{1}{2}, \frac{1}{2}) 0 0 0 0}.
\end{align}

\section{Details on the diagonalization}
\label{sec:details-diagonalization}

In this appendix, we provide details of the diagonalization procedure outlined in Section \ref{sec:mass-matrix}.

\subsection{Complicated bosons}
\label{sec:details-diagonalization-bosons}

In~\eqref{eq:Scomplicated} we have written the mass terms for the complicated bosons, i.e.\ those for which color and flavor degrees of freedom mix.
As stated in Section~\ref{sec:complicated-bosons}, the key observation is that we can diagonalize this mass term by starting with the $5 \times 5$ block for which we can rewrite the mixing term as
\begin{align}
    \frac{1}{2} S_{ij} L_{ij} = \frac{1}{2} \sum_{1\leq i < j\leq 5} \left[ (J_{ij})^2 - (L_{ij})^2 - (S_{ij})^2 \right].
\end{align}
We thus have to find the eigenstates of the total angular momentum operator $J_{ij} = L_{ij} + S_{ij}$.
Concretely, this works as follows.

The matrices $S_{ij}$ form the fundamental representation of \so{5},\footnote{In our conventions, $S_{jk}$ contains a $-i$ at position $(jk)$ and an $i$ at position $(kj)$.} and 
we bring them into canonical form by transforming the five complicated scalars as
\begin{align}
    \label{eq:scalars-flavor-transformation}
    \begin{pmatrix} \phi_1 \\ \phi_2 \\ \phi_3 \\ \phi_4 \\ \phi_5 \end{pmatrix}
    \to
    \frac{1}{\sqrt{2}}
    \begin{pmatrix}
        -i &  0 & 0 & -i & 0 \\
         1 &  0 & 0 & -1 & 0 \\
         0 & -i & i &  0 & 0 \\
         0 &  1 & 1 &  0 & 0 \\
         0 &  0 & 0 &  0 & i \sqrt{2}
    \end{pmatrix}^\dag
    \begin{pmatrix} \phi_1 \\ \phi_2 \\ \phi_3 \\ \phi_4 \\ \phi_5 \end{pmatrix}
    \equiv
    \begin{pmatrix}
        C_{++} \\ C_{+-} \\ C_{-+} \\ C_{--} \\ C_{00}
    \end{pmatrix}
    = \sum_{\alpha_1,\alpha_2} C_{\alpha_1,\alpha_2} \hat e_{\alpha_1,\alpha_2}.
\end{align}
The fields $C_{\alpha_1, \alpha_2}$ are the five components of the $(\tfrac{1}{2},\tfrac{1}{2})$ representation of \so{5}. 
In particular, we use the notation $C_{\alpha_1, \alpha_2} \equiv C_{\bS}$, where $\bS = (\tfrac{1}{2},\tfrac{1}{2}) \, |\alpha_1| |\alpha_2| \, \alpha_1 \alpha_2$, 
to make manifest that $C_{\alpha_1,\alpha_2}$ has magnetic quantum numbers $\alpha_1$ and $\alpha_2$ with respect to the $\su{2} \times \su{2}$ subalgebra of \so{5}.\footnote{Note that the subscripts $+$, $-$  and $0$ on the fields $C$ denote half-integers, e.g.\ $C_{+-}$ has $\alpha_1 = \tfrac{1}{2}$ and $\alpha_2 = -\tfrac{1}{2}$.}
These fields are now expanded in terms of \so{5} fuzzy spherical harmonics and we denote the components by $(C_{\bS})_{\bL}$.
Finally, the $\hat e_{\alpha_1,\alpha_2}$ in~\eqref{eq:scalars-flavor-transformation} are five-dimensional unit vectors, for example $\hat e_{++} = (1,0,0,0,0)$, and so on.

It is clear that the $(C_{\bS})_{\bL}$ transform as the product representation $(L_1,L_2) \otimes (\frac12,\frac12)$.
However, we are interested in fields that are diagonal with respect to the total angular momentum $J_{ij}$, and so will belong to the representations~\eqref{eq:5_times_irrep}. 
In particular, we will denote by $B_{\alpha_1,\alpha_2}$ the diagonal fields in the $(J_1,J_2) = (L_1 + \alpha_1, L_2 + \alpha_2)$ representation.
All the states in this total angular momentum representation are labelled by distinct values of $\bJ = (J_1, J_2) \, j_1 j_2 \, m_1 m_2$.
As familiar from quantum mechanics, the explicit change of basis is
\begin{align}
    \label{eq:fields-B-clebsch-gordan}
    (B_{\alpha_1,\alpha_2})_{\bJ} = \sum_{\bL, \bS} \langle \bL; \bS | \bJ \rangle (C_{\bS})_{\bL},
\end{align}
where $\langle \bL; \bS | \bJ \rangle$ are the Clebsch-Gordan coefficients for coupling the \so{5} states labeled by $\bL$ and $\bS$ to $\bJ$.
For the present case, i.e.\ the coupling of the fundamental of \so{5} with an arbitrary state in the irrep $(L_1, L_2)$, the coefficients can be found in~\cite{hecht1965}; see also Appendix~\ref{sec:action-generators} for more details.

The fields $(B_{\alpha_1,\alpha_2})_{\bJ}$ will have some corresponding basis elements $\hat Y^{\alpha_1,\alpha_2}_\bJ$, which are defined implicitly from
\begin{align}
 \sum_{\bL,\bS} (C_{\bS})_\bL \, \hat Y_\bL \otimes \hat e_{\bS}
 = 
 \sum_{\alpha_1,\alpha_2} \sum_\bJ 
 (B_{\alpha_1,\alpha_2})^{\ph \dag}_\bJ \, \hat Y_\bJ^{\alpha_1,\alpha_2}.
\end{align}
Having obtained eigenstates of the $5 \times 5$ block, it remains to see how they transform under the action of $\sum_{i=1}^5R_i^\dagger L_{i6}$, the $1 \times 5$ block in~\eqref{eq:Scomplicated}.
One can compute that
\begin{align}
    \left(\sum_{i=1}^5 R_i^\dagger L_{i6} \right) \hat{Y}_{\bJ}^{\alpha_1, \alpha_2} 
    = T^{P_1,P_2,P_3}_{J_1-\alpha_1, J_2-\alpha_2; J_1, J_2} \, \hat{Y}_{\bJ}.
\end{align}
The right-hand side of this equation is proportional to the \so{5} state $\hat{Y}_{\bJ}$ with a constant of proportionality $T$ that only depends on the irrep $(J_1, J_2)$ and $(\alpha_1, \alpha_2)$, \emph{not} on all quantum numbers contained in $\bJ$.
In fact, the $T$'s are certain reduced matrix elements of \so{6} generators; for more details, see Appendix~\ref{sec:action-generators}.
Their value also depends on which \so{6} representation the fields transform as and we will have to distinguish between the adjoint block with \so{6} irrep $(L_1 + L_2, L_1 + L_2, 0)$ and the off-diagonal block with $(\tfrac{n}{2}, \tfrac{n}{2}, \tfrac{n}{2})$.

Let us start with the adjoint block, in which case it turns out that the reduced matrix elements $T$ vanish if $(J_1, J_2) \in \{ (L_1 + \tfrac{1}{2}, L_2 + \tfrac{1}{2}), (L_1, L_2), (L_1 - \tfrac{1}{2}, L_2 - \tfrac{1}{2}) \}$.
More explicitly, we get
\begin{align}
\begin{split}
 \left(\sum_{i=1}^5 R_i^\dagger L_{i6} \right) \hat{Y}_{\bJ}^{++} &=
 \left(\sum_{i=1}^5 R_i^\dagger L_{i6} \right) \hat{Y}_{\bJ}^{--} =
 \left(\sum_{i=1}^5 R_i^\dagger L_{i6} \right) \hat{Y}_{\bJ}^{00} = 0, \\
 \left(\sum_{i=1}^5 R_i^\dagger L_{i6} \right) \hat{Y}_{\bJ}^{\pm\mp} &\equiv
 T^{\pm\mp} \hat{Y}_{\bJ},
\end{split}
\end{align}
where the coefficients $T^{\pm\mp}$ take the following values:
\begin{align}
  \label{eq:tpm-tmp-explicit}
\begin{split}
 T^{+-} 
 =  \sqrt{2} \sqrt{
    \frac{\left(2 J_1+1\right) \left(J_1-J_2\right) \left(J_2+1\right)}{2 J_1-2 J_2+1}
 }, \quad
 T^{-+}
 = -\sqrt{2} \sqrt{
    \frac{\left(2 J_1+3\right) \left(J_1-J_2+1\right) J_2}{2 J_1-2 J_2+1}
 }.
\end{split}
\end{align}
We now write the vector of complicated fields as
\begin{align}
    C =
    \begin{pmatrix}
        \sum_{\alpha_1, \alpha_2, \bJ} (B_{\alpha_1, \alpha_2})_{\bJ} \hat{Y}_{\bJ}^{\alpha_1, \alpha_2} \\
        \sum_{\bL} (A_3)_{\bL} \hat{Y}_{\bL}
    \end{pmatrix},
\end{align}
and insert into the mass term~\eqref{eq:Scomplicated}.
The mass term then becomes
\begin{align}
    \begin{split}
        &
        \hat m_{++}^2 (B_{++})^\dagger_{\bJ} (B_{++})_{\bJ} +
        \hat m_{--}^2 (B_{--})_{\bJ}^\dagger (B_{--})_{\bJ} +
        \hat m_{00}^2 (B_{00})^\dagger_{\bJ} (B_{00})_{\bJ} \\
        &+
        \begin{pmatrix}
            (B_{+-})_{\bJ}^\dagger & (B_{-+})_{\bJ}^\dagger & (A_3)_{\bJ}^\dagger
        \end{pmatrix}
        \begin{pmatrix}
            \measyh^2 + 2 & 0 & -\sqrt{2} T^{+-} \\
            0 & \measyh^2 + 2 & -\sqrt{2} T^{-+} \\
            -\sqrt{2} T^{+-} & -\sqrt{2} T^{-+} & \measyh^2
        \end{pmatrix}
        \begin{pmatrix}
            (B_{+-})_{\bJ} \\ (B_{-+})_{\bJ} \\ (A_3)_{\bJ}
        \end{pmatrix}.
    \end{split}
\end{align}
As pointed out above, the reduced Clebsch-Gordan coefficients only depend on the \so{5} and \so{6} irreps, not any other quantum numbers.
We can therefore simply diagonalize the remaining $3 \times 3$ matrix; the fields that achieve this diagonalization are given by
\begin{align}
    D_0 &= \frac{-1}{\sqrt{2 \measyh^2}} \left(T^{-+} B_{+-} - T^{+-} B_{-+} \right), \\
    D_\mp &= \frac{\pm 1 + \sqrt{4 \measyh^2 + 1}}{\sqrt{2 N_\pm}} A_3 \pm \frac{1}{\sqrt{N_\pm}} \left(T^{+-} B_{+-} + T^{-+} B_{-+} \right).
\end{align}
The eigenvalues are listed in Table~\ref{tab:masses-6x6}.

The diagonalization for the off-diagonal block proceeds similarly.
In this case the reduced matrix elements are non-zero only if $(J_1, J_2) = (L_1, L_2)$, resulting in a $2 \times 2$ matrix that has to be diagonalized in the final step.
The mass term becomes diagonal in terms of the fields $B_{++}$, $B_{-+}$ and
\begin{align}
    D_{\pm} = 
    \pm \sqrt{ \frac{1}{2} \pm \frac{1}{2 \sqrt{4\measy^2 + 1}} } B_{00}
    + \sqrt{ \frac{1}{2} \mp \frac{1}{2 \sqrt{4\measy^2 + 1}} } A_3.
\end{align}
The eigenvalues are listed in Table~\ref{tab:masses-6x6-offdiagonal}.

\subsection{Fermions}
\label{sec:details-diagonalization-fermions}

The mass term for the fermions as written in~\eqref{eq:mass-term-fermions-inserted-so5} is
\begin{align}
    \tr(\bar{\psi}_{\alpha} \mathcal{C}_{\alpha \beta} (P_L \psi_\beta) + \bar{\psi}_{\alpha} \mathcal{C}^\dagger_{\alpha \beta} (P_R \psi_\beta)),
\end{align}
where $P_L$ and $P_R$ are the chiral projectors.
The components of the matrix $\mathcal{C}_{\alpha \beta}$ are
\begin{align}
    \mathcal{C}_{\alpha \beta}
    =
    -\frac{1}{\sqrt{2}} \sum_{i = 1}^{5} (C_{i})_{\alpha \beta} L_{i6},
\end{align}
where the $(C_{i})_{\alpha \beta}$ were defined in \eqref{eq:def-dimred-g-matrices}.
One can show that $\mathcal{C}^\dagger \mathcal{C} \neq \mathcal{C} \mathcal{C}^\dagger$; thus, we cannot diagonalize $\mathcal{C}$ with a unitary transformation.
We will now follow a standard procedure to diagonalize a fermionic mass matrix used e.g.\ also in the standard model; see for example~\cite{Burgess:2007zi}.

We begin by finding the eigenvectors of $\mathcal{C}^\dagger \mathcal{C} = \tfrac{1}{2} (\sum_{i=1}^{5} \left(L_{i6}\right)^2 - \sum_{i,j=1}^{5}\tilde{S}_{ij} L_{ij}$).
The $4 \times 4$ matrices $\tilde{S}_{ij}$ form the four-dimensional representation of $\so{5}$; thus, $\mathcal{C}^\dagger \mathcal{C}$ is diagonalized by coupling a general \so{5} representation $(L_1, L_2)$ with $(\tfrac{1}{2}, 0)$.
As it was the case for the complicated bosons, we start by bringing the matrices $\tilde{S}_{ij}$ into canonical form with the transformation
\begin{align}
    \begin{pmatrix}
        \psi_1 \\ \psi_2 \\ \psi_3 \\ \psi_4
    \end{pmatrix}
    \to
    \frac{1}{2}
    \begin{pmatrix}
        1 & -i & -1 & i \\
        -i & 1 & i & -1 \\
        -1 & -i & -1 & -i \\
        i & 1 & i & 1
    \end{pmatrix}^\dagger
    \begin{pmatrix}
        \psi_1 \\ \psi_2 \\ \psi_3 \\ \psi_4
    \end{pmatrix}
    \equiv
    \begin{pmatrix}
        \tilde{C}_{+0} \\ \tilde{C}_{-0} \\ \tilde{C}_{0+} \\ \tilde{C}_{0-}
    \end{pmatrix}.
\end{align}
Here the fields $\tilde{C}_{\alpha_1 \alpha_2} \equiv (\tilde{C}_{\bS})_\bJ$ have well defined orbital and angular momentum.
Now the eigenvectors are found in terms of Clebsch-Gordan coefficients:
\begin{align}
    \hat{Y}_{\bJ}^{(L_1, L_2)} = \sum_{\bL, \bS} \, \langle \bL; \bS | \bJ \rangle \, \hat{Y}_{\bL} \otimes \hat{e}_{\bS}.
\end{align}
This concludes the diagonalization of $\mathcal{C}^\dagger \mathcal{C}$.

Now we will use the basis of eigenvectors of $\mathcal{C}^\dag \mathcal C$ to build a basis of eigenvectors of $\mathcal C$.
For the fields in the adjoint block, after a long calculation one can find how $\mathcal{C}$ acts on the four eigenvectors:
\begin{align}
    \mathcal{C} \, \hat{Y}_{\bJ}^{(J_1 \pm \frac{1}{2}, J_2)}
    &=
    \chi_1(\bJ) \, m_{\pm 0}(J_1, J_2) \, \left(\hat{Y}_{\bJ_{\mathrm{r}}}^{(J_1, J_2 \pm \frac{1}{2})}\right)^\star, \\
    \mathcal{C} \, \hat{Y}_{\bJ}^{(J_1, J_2 \pm \frac{1}{2})}
    &=
    \chi_2(\bJ) \, m_{0 \pm}(J_1, J_2) \, \left(\hat{Y}_{\bJ_{\mathrm{r}}}^{(J_1 \pm \frac{1}{2}, J_2)}\right)^\star,
\end{align}
with the `reversed' total angular momentum $\bJ_{\mathrm{r}} \equiv (J_1, J_2) \, j_2 j_1 \, m_2 m_1$ and some phase factors $\chi_1(\bJ)$ and $\chi_2(\bJ)$.
It turns out that $m_{\pm 0}$ and $m_{0 \pm}$ are the same when written in terms of $J_1$ and $J_2$ so that we can obtain eigenvectors of $\mathcal{C}$ by essentially adding the two previous equations and taking care of the phase factors.
After the dust has settled, the eigenvectors of $\mathcal{C}$ turn out to be
\begin{align}
\label{eq:extra-step}
    \hat{Y}_{\bJ}^{\alpha \beta}
    =
    \frac{\chi(\bJ; \alpha, \beta) }{\sqrt{2}} \left[ \hat{Y}_{\bJ}^{(J_1 + \frac{\alpha}{2}, J_2)} + \beta \hat{Y}_{\bJ_{\mathrm{r}}}^{(J_1, J_2 + \frac{\alpha}{2})} \right],
\end{align}
for the four combinations of $\alpha, \beta \in \{-1, +1\}$ and the phase
\begin{align}
    \chi(\bJ; \alpha, \beta) = (-1)^{-\frac{1}{2} (2 J_1 + m_1 + m_2 + \alpha + \frac{1}{2})} i^{\frac{1 - \beta}{2}}.
\end{align}
The fermions in the action can now be expanded in this basis and the mass term becomes diagonal in terms of component fields which we call $(\Bfermion_{\alpha \beta})_{\bJ}$, and which are related to $(\tilde{C}_{\bS})_\bJ$ by 
\begin{align}
    \sum_{\alpha,\beta} \sum_{\bJ} \, (\Bfermion_{\alpha \beta})_{\bJ} \, \hat{Y}_{\bJ}^{\alpha \beta}
    =
    \sum_{\bS, \bL} \, (\tilde C_\bS)_{\bL} \, \hat{Y}_{\bL} \otimes \hat{e}_{\bS}.
\end{align}

One can diagonalize the fields in the off-diagonal block in a similar fashion, the only difference being that different orbital angular momentum representations are not mixed with each other.
There is still mixing between $\bJ$ and $\bJ_r$, which can be diagonalized easily with an extra step similar to~\eqref{eq:extra-step}.

\section{Details on the propagators}
\label{sec:propagator-details}

In this appendix, we provide further details on the derivation of the propagators presented in Section \ref{sec:propagators}.
In particular, we give the explicit formulas for the coefficients $\hat f$ that do not appear in the main text.

The fields in which the mass matrix for the bosons becomes diagonal are $B_{\pm, \pm}$, $B_{0,0}$, $D_{\pm}$ and $D_{0}$.
The propagators between them are simply
\begin{align}
    \langle [B_{++}]_{\bL} [B_{++}]_{\bL'}^\dagger \rangle &= \delta_{\bL, \bL'} K^{\hat{m}_{++}^2},
\end{align}
and similarly for $B_{--}$, $B_{00}$, $D_{\pm}$ and $D_0$.
In order to invert the Clebsch-Gordan procedure, we have to express the non-diagonal fields $B_{\pm, \mp}$ and $A_{3}$ in terms of the diagonal fields.
This is achieved by
\begin{align}
   \label{eq:b-dp-dm}
    B_{\pm, \mp} &=
    \mp \frac{T^{\mp, \pm}}{\sqrt{2 \measyh^2}} D_0
    - T^{\pm, \mp} \left( \frac{D_+}{\sqrt{N_-}} - \frac{D_-}{\sqrt{N_+}} \right), \\
    \label{eq:a3-dp-dm}
    A_3 &=
    \frac{-1 + \sqrt{4 \measyh^2 + 1}}{\sqrt{2 N_-}} D_+ + \frac{1 + \sqrt{4 \measyh^2 + 1}}{\sqrt{2 N_+}} D_-.
\end{align}
From~\eqref{eq:a3-dp-dm} it is immediate to obtain the propagator $\langle A_3 \, A_3^\dag \rangle$, see~\eqref{eq:propagator-A3-A3-main} in the main text.
Similarly, for the propagator $\langle \phi_i \, A_3^\dag \rangle$ the two fields couple through propagators $\langle B_{\pm,\mp} \, A_3^\dag \rangle$. 
It is therefore natural to introduce the following function
\begin{align}
    \hat{f}^{\,\phi A}(L_1, L_2)
    &=
    \frac{-1 + \sqrt{4 \measyh^2 + 1}}{\sqrt{2} N_-} K^{\hat{m}_{+}^2} -
    \frac{1 + \sqrt{4 \measyh^2 + 1}}{\sqrt{2} N_+} K^{\hat{m}_{-}^2},
\end{align}
which captures such contributions.\footnote{The prefactor $T^{\pm\mp}$ that would naively appear gets absorbed in the matrix element of $L_{i6}$, as one can see by doing the calculation of the propagators carefully. A similar prefactor will also get absorbed by the matrix elements of the generators in~\eqref{eq:fopp}.}

The situations is more complicated for the propagators $\langle \phi_i \, \phi_j^\dag \rangle$, because there are several possible contributions.
The first one comes from the propagator $\langle B_{+,-} \, B_{-,+}^\dag \rangle$, and it is captured by the function
\begin{align}
\label{eq:fopp}
 \hat{f}^{\,\mathrm{opp}}(L_1, L_2)
 &=
 \frac{1}{2} \left(- \frac{K^{\hat{m}_0^2}}{2 \measyh^2} + \frac{K^{\hat{m}_{+}^2}}{N_-} + \frac{K^{\hat{m}_{-}^2}}{N_+} \right).
\end{align}
The other contributions come from propagators between identical $B$ fields $\langle B_{\alpha,\beta} \, B_{\alpha,\beta}^\dag \rangle$, and we will encode them in the functions $h_{\alpha,\beta}$.
These functions are particularly simple for the $B$ fields that are diagonal after the Clebsch-Gordan decomposition
\begin{align}
  h_{\pm, \pm}(L_1, L_2) = K^{\hat{m}_{\pm \pm}^2}, \qquad
  h_{0, 0}(L_1, L_2) = K^{\hat{m}_{00}^2}.
\end{align}
For the fields $B_{\pm,\mp}$, we can read off the corresponding contribution from~\eqref{eq:b-dp-dm}, namely
\begin{align}
  \begin{split}
    h_{\pm, \mp}(L_1, L_2) &= \frac{(T^{\mp \pm})^2}{2 \measyh^2} K^{\hat{m}_0^2} + (T^{\pm \mp})^2 \left(\frac{K^{\hat{m}_{+}^2}}{N_-} + \frac{K^{\hat{m}_{-}^2}}{N_+} \right).
  \end{split}
\end{align}
Note that here the $T^{\pm,\mp}$ given in~\eqref{eq:tpm-tmp-explicit} are to be evaluated at $(L_1, L_2)$, i.e.\ one has to replace $(J_1, J_2) \rightarrow (L_1, L_2)$.

The functions $\hat f$ and $h_{\alpha\beta}$ we just defined are the building blocks of the final propagators.
In order to obtain the full expressions, we start with a certain propagator, and expand it using~\eqref{eq:fields-B-clebsch-gordan} and~\eqref{eq:scalars-flavor-transformation}, and then evaluate the propagators of $B$ fields and $A_3$ in the way we just described.
The result will be a complicated combination of products of Clebsch-Gordan coefficients and the functions $\hat f$ and $h_{\alpha\beta}$.
These expressions can always be rewritten in terms of matrix elements of \so{6} generators\footnote{In practice, it is easiest to make an ansatz for the propagators and if the coefficients can be fixed for all possible combination, then the ansatz is correct.} to obtain the form presented in Section~\ref{sec:prop-central-block}.

In~\eqref{eq:prop-bosons-complicated} we have written the propagators between the scalars in terms of the functions $\hat{f}^{\,\mathrm{sing}}$, $\hat{f}^{\,\mathrm{cub}}$, $\hat{f}^{\,\mathrm{lin}}$, $\hat{f}^{\,\mathrm{sym}}_{5}$, $\hat{f}^{\,\mathrm{sym}}_{6}$ and $\hat{f}^{\,\mathrm{opp}}$ that are linear combinations of propagators between mass eigenstates.
To write them in a more compact way, we define
\begin{align}
 Z_{\alpha, \beta}(L_1, L_2) \equiv \frac{1}{2} \left( C_5(L_1 + \tfrac{\alpha}{2}, L_2 + \tfrac{\beta}{2}) - C_5(L_1, L_2) \right),
\end{align}
and
\begin{align}
 D_{\alpha, \beta}(L_1, L_2) \equiv
 \begin{cases}
  i^{\alpha - \beta} 2 (L_1 + 1) (2 L_2 + 1) Z_{\alpha, \beta} (2 Z_{\alpha, \beta} - 1) & (\alpha, \beta) \neq (0, 0)\,, \\
  \prod_{(\gamma, \delta) \neq (0, 0)} Z_{\gamma, \delta} & (\alpha, \beta) = (0, 0)\,.
 \end{cases}
\end{align}
The indices $(\alpha,\beta)$ run over the five values $(\pm 1, \pm 1)$ and $(0,0)$.
After a complicated calculation, on can see that $\hat f$ are given by\footnote{In the following equations $Z_{\alpha,\beta}$, $D_{\alpha,\beta}$ and $C_5$ are always evaluated at $(L_1, L_2)$ unless noted otherwise.}
\begin{align}
\begin{split}
 \hat{f}^{\,\mathrm{sing}}(L_1, L_2)
 =
 \sum_{(\alpha, \beta)} \frac{1}{2 D_{\alpha, \beta}}
 \bigg[ &-2 Z_{\alpha, \beta}^2 \left( 1 + C_5 - Z_{\alpha, \beta}^2 \right)- C_5 - 2 (Z_{+,-} Z_{-,+})^2\\
 & - 2 (1 + C_5) Z_{+,-} Z_{-,+}  \bigg] \, h_{\alpha, \beta}(L_1 + \tfrac{\alpha}{2}, L_2 + \tfrac{\beta}{2})\,,
\end{split}
\end{align}
and
\begin{align}
 \hat{f}^{\,\mathrm{lin}}(L_1, L_2)
 &=
 \sum_{(\alpha, \beta)} \frac{i}{4 D_{\alpha, \beta}}
 (2 Z_{\alpha, \beta} + 1) (2 C_5 - 2 Z_{\alpha, \beta}^2 - 3) \, h_{\alpha, \beta}(L_1 + \tfrac{\alpha}{2}, L_2 + \tfrac{\beta}{2})\,, \\
 \hat{f}^{\,\mathrm{cub}}(L_1, L_2)
 &=
 \sum_{(\alpha, \beta)} \frac{-i}{4 D_{\alpha, \beta}}
 (2 Z_{\alpha, \beta} + 1) \, h_{\alpha, \beta}(L_1 + \tfrac{\alpha}{2}, L_2 + \tfrac{\beta}{2})\,, \\
 \hat{f}^{\,\mathrm{sym}}_{5}(L_1, L_2)
 &=
 \sum_{(\alpha, \beta)} \frac{-1}{2 D_{\alpha, \beta}}
 \left( \frac{1}{2} + Z_{+,-} Z_{-,+} + Z_{\alpha, \beta}^2 \right) \, h_{\alpha, \beta}(L_1 + \tfrac{\alpha}{2}, L_2 + \tfrac{\beta}{2})\,, \\
 \hat{f}^{\,\mathrm{sym}}_{6}(L_1, L_2)
 &=
 \sum_{(\alpha, \beta)} \frac{-1}{4 D_{\alpha, \beta}}
 (2 Z_{+,-} + 1) (2 Z_{-,+} + 1) \, h_{\alpha, \beta}(L_1 + \tfrac{\alpha}{2}, L_2 + \tfrac{\beta}{2})\,.
\end{align}
As the reader can observe, the functions $D_{\alpha\beta}$ and $Z_{\alpha\beta}$ allowed to compactly write the $\hat f$, but we do not think they have any physical meaning beyond this.

In order to obtain the fermionic propagators, we follow an identical procedure as described above.
We start with a given propagator, expand it following the steps described in the diagonalization, and then identify the result in terms of propagators of diagonal fields and matrix elements of \so6 generators.
The result is given by~\eqref{eq:ferm-prop-main} and~\eqref{eq:ferm-prop-main-g5}, where the explicit expressions for $\hat{f}_F$ are
\begin{align}
\begin{split}
 \hat{f}_F^{\,\mathrm{lin}}(L_1, L_2; L_1 + \tfrac12, L_2 - \tfrac12)
 & =
 \frac{\left(L_1+L_2+1\right) K_{F}^{m=\sqrt{2(L_1+1)(L_2+1)}}}
      {\sqrt{\left(2 L_1+3\right) \left(2 L_2+1\right)} \left(2 L_1+2 L_2+3\right)} \\
 & \phantom{{}={}}+
 \frac{\left(L_1+L_2+2\right) K_{F}^{m=\sqrt{2 L_2 (L_1+1)}}}
      {2 \sqrt{\left(L_1+1\right) L_2} \left(2 L_1+2 L_2+3\right)}, \\[0.5em]
 \hat{f}_F^{\,\mathrm{lin}}(L_1, L_2; L_1 - \tfrac12, L_2 + \tfrac12)
 & = 
 \frac{\left(L_1+L_2+1\right) K_{F}^{m=\sqrt{2(L_1+1)(L_2+1)}}}
      {2 \sqrt{\left(L_1+1\right) \left(L_2+1\right)} \left(2 L_1+2 L_2+3\right)} \\
 & \phantom{{}={}}+
 \frac{\left(L_1+L_2+2\right) K_{F}^{m=\sqrt{2 L_2 (L_1+1)}}}
      {\sqrt{\left(2 L_1+1\right) \left(2 L_2+1\right)} \left(2 L_1+2 L_2+3\right)},
\end{split}
\end{align}
and
\begin{align}
\begin{split}
 \hat{f}_F^{\,\mathrm{cub}}(L_1, L_2; L_1 + \tfrac12, L_2 - \tfrac12)
 & = 
 \frac{3 K_{F}^{m=\sqrt{2(L_1+1)(L_2+1)}}}
      {\sqrt{\left(2 L_1+3\right) \left(2 L_2+1\right)} \left(2 L_1+2 L_2+3\right)} \\
 & \phantom{{}={}}-
 \frac{3 K_{F}^{m=\sqrt{2 L_2 (L_1+1)}}}
      {2 \sqrt{\left(L_1+1\right) L_2} \left(2 L_1+2 L_2+3\right)}, \\[0.5em]
 \hat{f}_F^{\,\mathrm{cub}}(L_1, L_2; L_1 - \tfrac12, L_2 + \tfrac12)
 & = 
 \frac{3 K_{F}^{m=\sqrt{2(L_1+1)(L_2+1)}}}
      {2 \sqrt{\left(L_1+1\right) \left(L_2+1\right)} \left(2 L_1+2 L_2+3\right)} \\
 & \phantom{{}={}}-
 \frac{3 K_{F}^{m=\sqrt{2 L_2 (L_1+1)}}}
      {\sqrt{\left(2 L_1+1\right) \left(2 L_2+1\right)} \left(2 L_1+2 L_2+3\right)}.
\end{split}
\end{align}

\section{Effective vertex}
\label{sec:effective-vertex}

In this appendix, we will give some extra details on how to compute the effective vertex.
We remind the reader that we started with the $\mathcal N = 4$ SYM action, and we expanded around a classical solution $\phi_i = \pcl_i + \pt_i$. 
This gives rise to a number of cubic interaction vertices:
\begin{align}
 \label{eq:action-cubic-vertices}
 \begin{aligned}
 S_3 
  & = \frac{2}{\gym^2} \int \mathrm{d}^4 x \; \tr \Bigg(
    i [ A^\mu, A^\nu ] \partial_\mu A_\nu  +
      \pt_i [\pt_j , [ \pcl_i, \pt_j ]]   +
    i [ A^\mu, \pt_i ] \partial_\mu \pt_i  +
      \pt_i [ A^\mu, [ \pcl_i , A_\mu ]] \\ & +
    \frac{1}{2} \bar{\psi} \gamma^\mu [ A_\mu, \psi ]  +
    \frac{1}{2} \sum_{i=1}^3 \bar{\psi} C_i [ \pt_i, \psi ] +
    \frac{1}{2} \sum_{i=4}^6 \bar{\psi} C_i [ \pt_i, \gamma_5 \psi ] +
    i (\partial_\mu \bar c) [ A_\mu, c ] -
    \bar{c} [ \pcl_i [ \pt_i, c ]]
  \Bigg).
 \end{aligned}
\end{align}
These are the only vertices that can contribute to the computation of the effective vertex.
The following calculation proceeds in exactly the same manner as that of~\cite{Buhl-Mortensen:2016jqo,Grau:2018keb}.
We will only write the contractions that contribute, all other possible Wick contractions being zero.

There is one contribution from the ghost fields, which behave simply as easy scalars
\begin{align}
 \contraction{- \tr \Big(}{\bar c}{[ \pcl_i, [\tilde \phi_i,}{c}
 - \tr \left( \bar c [ \pcl_i, [\tilde \phi_i,c]] \right)
 & = \frac{\sqrt{2} N}{y_3} K^{\measy^2} \tr \left(\pt_i G_{i6}\right).
\end{align}
Only two contractions survive in the vertex that couples two scalars with the gauge field%
\footnote{In the second contraction, we can use (D.21) from~\cite{Buhl-Mortensen:2016jqo}, since we have
\begin{align}
 \nu_- = \sqrt{m_-^2 + \frac{1}{4}} = \nu_{\mathrm{easy}} - 1, \quad
 \nu_+ = \sqrt{m_+^2 + \frac{1}{4}} = \nu_{\mathrm{easy}} + 1,
\end{align}
for both the fields in the diagonal and in the off-diagonal blocks, and the propagator $\hat{K}^{A\phi}$ has the desired form $K^{\nu - 1} - K^{\nu + 1}$.}
\begin{align}
 \contraction{\tr \big( i [}{A^\mu}{,}{\tilde \phi_i}
 \tr \big( i [A^\mu, \tilde \phi_i] \partial_\mu \tilde \phi_i \big)
 +
 \contraction{\tr \big( i [}{A^\mu}{,\tilde \phi_i] \partial_\mu}{\tilde \phi_i}
 \tr \big( i [A^\mu, \tilde \phi_i] \partial_\mu \tilde \phi_i \big)
 = + 6 i N \partial_3 f^{A\phi} \tr \left( \pt_i \, G_{i6} \right),
\end{align}
where
\begin{align}
    f^{A \phi} = \frac{-i}{2 \sqrt{n (n + 4) + 2}} \left(K^{m_-^2} - K^{m_+^2}\right).
\end{align}
For the vertex that couples three scalars, all possible Wick contractions contribute
\begin{subequations}
 \begin{align}
  \contraction{\tr \big( \pt_i [ }{\pt_j}{, [\pcl_i, }{\pt_j}
  \tr \big( \pt_i [ \pt_j, [\pcl_i, \pt_j]] \big)
  &=
  - \frac{\sqrt{2} N}{y_3} \Big[ 
      5 f^{\mathrm{sing}}
    + n(n + 4) f^{\mathrm{prod}}
    + K^{\measy^2}
   \Big] \tr \left( \pt_i G_{i6} \right), \\
  \contraction{\tr \big( }{\pt_i}{[ \pt_j, [\pcl_i, }{\pt_j}
  \tr \big( \pt_i [ \pt_j, [\pcl_i, \pt_j]] \big)
  & = \frac{\sqrt{2} N}{y_3} \Big[ 
      f^{\mathrm{sing}}
    + 2i f^{\mathrm{lin}}  
    + \big[ n(n + 4) - 8 \big] f^{\mathrm{prod}}
   \Big] \tr \left( \pt_i G_{i6} \right), \\
  \contraction{\tr \big(}{\pt_i}{[}{\pt_j}
  \tr \big( \pt_i [ \pt_j, [\pcl_i, \pt_j]] \big)
   & = \frac{4\sqrt{2} N}{y_3} \Big[ 
      i f^{\mathrm{lin}}  
    - 2 f^{\mathrm{prod}}
   \Big] \tr \left( \pt_i G_{i6} \right).
\end{align}
\end{subequations}
The regularization procedure becomes important when we consider the vertex that couples two gauge fields and a scalar.
We work in dimensional reduction~\cite{Siegel:1979wq,Capper:1979ns}  with ${d = 3 - 2 \epsilon}$ space dimensions, hence $n_{\mathrm{A,easy}} = 3 - 2 \epsilon$ and we should  add $2 \epsilon$ scalars to the action that behave exactly as the easy components of the gauge field. The choice of this regularization procedure is motivated by the fact that it is supersymmetry preserving and hence compatible with the symmetries of the bulk ${\cal N}=4$ SYM theory which we must recover far
from the domain wall, cf.\ the discussion in~\cite{Buhl-Mortensen:2016jqo,Grau:2018keb}.
In total, we get
\begin{align}
\begin{split}
 \contraction{\tr \big( \pt_i [}{A^\mu}{, [\pcl_i, }{A_\mu}
 \tr \big( \pt_i [ A^\mu, [\pcl_i, A_\mu]] \big)
 & +
 \contraction{\tr \big( \pt_i [}{A^{2 \epsilon}}{, [\pcl_i, }{A_{2 \epsilon}}
 \tr \big( \pt_i [ A^{2 \epsilon}, [\pcl_i, A_{2 \epsilon}]] \big) \\
 & = - \frac{\sqrt{2} N}{y_3} \left(
     (n_{A,\mathrm{easy}} + 2\epsilon ) K^{\measy^2} 
     +  f^{AA} 
 \right) \tr \left(\pt_i G_{i6} \right),
\end{split}
\end{align}
where
\begin{align}
    f^{AA} = \frac{1}{2} \left[
        \left(1 + \frac{1}{\sqrt{4 \measy^2 + 1}} \right) K^{m_-^2} +
        \left(1 - \frac{1}{\sqrt{4 \measy^2 + 1}} \right) K^{m_+^2}
    \right].
\end{align}
Finally, we can also have fermions running in the loop, which contribute as
\begin{align}
   \frac{1}{2} \sum_{i=1}^3 
   (C_i)_{\alpha\beta} 
   \contraction{\tr \bigg(}{\bar \psi_\alpha}{[ \pt_i,}{\psi_\beta}
   \tr \bigg( \bar \psi_\alpha [ \pt_i, \psi_\beta] \bigg)
 + \frac{1}{2} \sum_{i=4}^5 
   (C_i)_{\alpha\beta} 
   \contraction{\tr \bigg(}{\bar \psi_\alpha}{[ \pt_i, \gamma_5}{\psi_\beta}
   \tr \bigg( \bar \psi_\alpha [ \pt_i, \gamma_5 \psi_\beta] \bigg)
 = 8 N \tr f_{F}^{2,+} \, \tr(\pt_i G_{i6}).
\end{align}

One can sum all the contributions above, and simplify the resulting expression using identities such as $\Psi(z + 1) = \Psi(z) + 1/z$.
The result that one obtains is~\eqref{eq:effective-vertex}, where one notices that the dependence on the regulator $\epsilon$ drops completely.

\section{Matrix elements and Clebsch-Gordan coefficients}
\label{sec:action-generators}

In this appendix, we describe how to compute matrix elements of $\so{6}$ generators acting on general representations and where to obtain the Clebsch-Gordan coefficients relevant for the calculations in this work.

\renewcommand*{\arraystretch}{1.5}
\begin{table}
\centering
\begin{tabular}{l l}
 \toprule
 Labels $\mathbf S$ &
 Tensor operator $T_{\mathbf S}$ \\
 \midrule
 $((1,0), 1, 0, 0, 0)$ & 
 $\frac{1}{2}( L_{12} + L_{34} )$ \\
 $((1,0), 1, 0, \pm 1, 0)$ & 
 $\frac{1}{2\sqrt{2}}( \mp (L_{14} + L_{23}) + i (L_{13} - L_{24}) )$ \\
 $((1,0), 0, 1, 0, 0)$ & 
 $\frac{1}{2}( L_{12} - L_{34} )$ \\
 $((1,0), 1, 0, \pm 1, 0)$ & 
 $\frac{1}{2\sqrt{2}}( \mp (L_{14} - L_{23}) - i (L_{13} + L_{24}) )$ \\
 $((1,0), \frac{1}{2}, \frac{1}{2}, \pm \frac{1}{2}, \pm \frac{1}{2})$ & 
 $\frac{1}{2}( \pm L_{25} - i L_{15})$ \\
 $((1,0), \frac{1}{2}, \frac{1}{2}, \pm \frac{1}{2}, \mp \frac{1}{2})$ & 
 $\frac{1}{2}( L_{45} \mp i L_{35})$ \\
 $((\frac{1}{2},\frac{1}{2}), 0, 0, 0, 0)$ &
 $-L_{56}$ \\
 $((\frac{1}{2},\frac{1}{2}), \frac{1}{2},\frac{1}{2}, \pm \frac{1}{2}, \pm\frac{1}{2})$ & 
 $\frac{1}{\sqrt{2}}( L_{16} \pm i L_{26} )$ \\
 $((\frac{1}{2},\frac{1}{2}), \frac{1}{2},\frac{1}{2}, \pm \frac{1}{2}, \mp\frac{1}{2})$ & 
 $\frac{1}{\sqrt{2}}( \pm L_{36} + i L_{46} )$ \\
 \bottomrule
\end{tabular}
\caption{Relation between the tensor operators of $\so{6}$ and the corresponding generators $L_{ij}$.}
\label{table:tensor-ops}
\end{table}
\renewcommand*{\arraystretch}{1.0}

In Table~\ref{table:tensor-ops} we map the generators $L_{ij}$ to the tensor operators $T_{\mathbf S}$, as the latter have much simpler matrix elements.
Notice how these tensor operators are labeled by a set of \so{5} quantum numbers $\mathbf S = (S_1,S_2), s_1, s_2, m_1, m_2$.
The tensor operators which transform in the ten-dimensional representation $(1,0)$ of $\so{5}$ only act on the \so{5} labels $\bL$.
The matrix elements are
\begin{align}
 \label{eq:mat-el-so5}
 \langle \mathbf L' | T_{\mathbf S} | \mathbf L \rangle 
 = \delta_{L_1,L_1'} \delta_{L_2,L_2'} 
   \sqrt{L_1(L_1+2) + L_2(L_2+1)}
   \langle \mathbf L; \mathbf S | \mathbf L' \rangle.
\end{align}
The square root is sometimes called a reduced matrix element or isoscalar factor, and the second term is an \so{5} Clebsch-Gordan coefficient from coupling $\mathbf L$ and $\mathbf S$.

On the other hand, the tensor operators which transform in the five-dimensional representation $(\frac12,\frac12)$ of $\so{5}$ will affect both the $\so{5}$ and \so{6} quantum numbers.
Therefore, we compute matrix elements of these operators with $\so{6}$ states with labels $\mathbf P = (P_1,P_2,P_3)\mathbf L$, where $\mathbf L$ are the labels of the \so{5} subgroup.
Then, the matrix elements are
\begin{align}
 \label{eq:mat-el-so6}
 \langle \mathbf P' | T_{\mathbf S} | \mathbf P \rangle 
 = \delta_{P_1,P_1'} \delta_{P_2,P_2'} \delta_{P_3,P_3'} \,
   T^{P_1,P_2,P_3}_{L_1,L_2;L_1',L_2'} \,
   \langle \mathbf L; \mathbf S | \mathbf L' \rangle.
\end{align}
As before, the matrix element is a product of a reduced matrix element $T^{P_1, P_2, P_3}_{L_1, L_2; L_1', L_2'}$ and an $\so{5}$ Clebsch-Gordan coefficient.

The reduced matrix elements $T^{P_1, P_2, P_3}_{L_1, L_2; L_1', L_2'}$ that appear in~\eqref{eq:mat-el-so6} are more complicated than those in~\eqref{eq:mat-el-so5} and we have derived them using the strategy described in~\cite{hecht1965}.
The main idea is the following.
On the one hand, a construction by Gel'fand and Tsetlin~\cite{gelfand1950} gives the matrix elements of \so{n} generators for any $n$.
On the other hand, these matrix elements factorize into \so{n - 1} Clebsch-Gordan coefficients and the reduced matrix elements that we are after.
This factorization is the content of the Wigner-Eckart theorem.
Since the relevant \so{5} Clebsch-Gordan coefficients are known, e.g.\ from~\cite{hecht1965}, one can construct the matrix elements for \so{6} and essentially compare the two expressions.
The missing factors are then the reduced matrix elements, which we present in Table~\ref{table:reduced-matrix-el}.

We have shown that with knowledge of certain \so{5} Clebsch-Gordan coefficients one can construct the matrix elements for any \so{6} generator. 
The \so{5} Clebsch-Gordan coefficients factorize as 
\begin{align}
\begin{split}
 \langle (L_1, L_2), \ell_1,\ell_2, m_{\ell1}, m_{\ell2}; 
         (S_1,&S_2), s_1 ,s_2   , m_{s1}, m_{s2} |
         (J_1, J_2), j_1   ,j_2   , m_{j1}, m_{j2} \rangle
  \\ 
  & = 
  \langle (L_1,L_2), \ell_1,\ell_2; 
          (S_1,S_2), s_1,s_2 ||
          (J_1,J_2), j_1,j_2 \rangle \\
  & \quad \times
    \langle \ell_1,m_{\ell1}; s_1, m_{s1} | j_1, m_{j1} \rangle
    \langle \ell_2,m_{\ell2}; s_2, m_{s2} | j_2, m_{j2} \rangle.
\end{split}
\end{align}
The double-barred coefficients are reduced \so{5} Clebsch-Gordan coefficients, while the other two terms are usual $\su{2}$ Clebsch-Gordan coefficients.
The reduced coefficients were computed in~\cite{hecht1965} for the cases $(S_1,S_2) = (\frac{1}{2},0), (\frac{1}{2},\frac{1}{2}), (1,0)$.\footnote{In the notation from~\cite{hecht1965} one has $J_m = L_1$, $\Lambda_m = L_2$, $\ell_1 = J$, and so on. Except for these minor notation differences, our conventions are identical to theirs, and one can directly extract the double-barred coefficients from the tables at the end of that paper.}
In order to make it easy for the interested reader to reproduce our results, we attach a \texttt{Mathematica} file with all the relevant \so{5} Clebsch-Gordan coefficients and the reduced matrix elements from Table~\ref{table:reduced-matrix-el}.
We are also happy to provide more details on request.

\begin{landscape}
  \begin{table}
  \centering
  \begin{adjustbox}{center}
  \begin{tabular}{l l}
  \toprule
  $(L_1',L_2')$ &
  $T^{P_1,P_2,P_3}_{L_1,L_2;L_1',L_2'}$ \\
  \midrule
  $(L_1-\frac{1}{2},L_2-\frac{1}{2})$ &
  $\displaystyle \left(
      \frac{(L_1 + L_2 - P_1 - 1) (L_1 + L_2 + P_1 + 3)
            (L_1 + L_2 - P_2) (L_1 + L_2 + P_2 + 2) (L_1 + L_2 - P_3 + 1) 
            (L_1 + L_2 + P_3 + 1)}
          {2 (2 L_1 + 1) L_2 (L_1 + L_2 + 1) (2 L_1 + 2 L_2 + 1)}
      \right)^{1/2}$ \\
  $(L_1-\frac{1}{2},L_2+\frac{1}{2})$ &
  $\displaystyle  -\left(
      \frac{(L_1 - L_2 - P_1 - 2) (L_1 - L_2 + P_1 + 2)
            (L_1 - L_2 - P_2 - 1) (L_1 - L_2 + P_2 + 1) 
            (L_1 - L_2 - P_3) (L_1 - L_2 + P_3)}
           {2 (2 L_1 + 1) (L_2 + 1) (L_1 - L_2) (2 L_1 - 2 L_2 - 1)}
  \right)^{1/2}$ \\
  $(L_1,L_2)$ &
  $\displaystyle  -\left(
      \frac{(P_1 + 2)^2 (P_2 + 1)^2 {P_3}^2}
           {(L_1- L_2) (L_1 - L_2 + 1) (L_1 + L_2 + 1) (L_1 + L_2 + 2)}
  \right)^{1/2}$ \\
  $(L_1+\frac{1}{2},L_2-\frac{1}{2})$ &
  $\displaystyle \left(
      \frac{(L_1 - L_2 - P_1 - 1) (L_1 - L_2 + P_1 + 3) 
            (L_1 - L_2 - P_2) (L_1 - L_2 + P_2 + 2) 
            (L_1 - L_2 - P_3 + 1) (L_1 - L_2 + P_3 + 1)}
           {2 (2 L_1 + 3) L_2 (L_1 - L_2 + 1) (2 L_1 - 2 L_2 + 3)}
  \right)^{1/2}
  $ \\
  $(L_1+\frac{1}{2},L_2+\frac{1}{2})$ &
  $\displaystyle  -\left(
      \frac{ (L_1 + L_2 - P_1) (L_1 + L_2 + P_1 + 4) 
             (L_1 + L_2 - P_2 + 1) (L_1 + L_2 + P_2 + 3) 
             (L_1 + L_2 - P_3 + 2) (L_1 + L_2 + P_3 + 2) }
           {2 (2 L_1 + 3) (L_2 + 1) (L_1 + L_2 + 2) (2 L_1 + 2 L_2 + 5)} 
  \right)^{1/2}$ \\
  \bottomrule
  \end{tabular}
  \end{adjustbox}
  \caption{Reduced matrix elements appearing in~\eqref{eq:mat-el-so6}.}
  \label{table:reduced-matrix-el}
  \end{table}
\end{landscape}

\providecommand{\href}[2]{#2}\begingroup\raggedright\endgroup

\end{document}